\documentclass[usedcolumn]{aa}
\usepackage{times}
\usepackage{mathptm}
\usepackage{ptmgreek}
\usepackage{epsf}
\newcommand{\etal}{{et al}\/.}
\begin{document}
\title{An all-sky optical
  catalogue of radio / X-ray sources}
\titlerunning{Optical catalogue of radio / X-ray sources}
\author{E. Flesch\inst{1} \and M.J.~Hardcastle\inst{2}}
\institute{P.O. Box 12520, Wellington, New Zealand (eric@flesch.org) \and
Department of Physics, University of Bristol, Tyndall
Avenue, Bristol BS8 1TL, UK (m.hardcastle@bristol.ac.uk)}
\date{Version of \today}

\abstract{We present an all-sky catalogue that aligns and overlays the
{\it ROSAT} HRI, RASS, PSPC and WGA X-ray catalogues and the NVSS,
FIRST and SUMSS radio catalogues onto the optical APM and USNO-A
catalogues. Objects presented are those APM/USNO-A optical objects
which are calculated with $\ge 40$\% confidence to be associated with
radio/X-ray detections, or which are identified as known QSOs, AGN or
BL Lacs, totalling 501,761 objects in all, including 48,285 QSOs and
21,498 double radio lobe detections. For each radio/X-ray associated
optical object we display the calculated percentage probabilities of
its being a QSO, galaxy, star, or erroneous radio/X-ray association,
plus any identification from the literature. The catalogue includes
86,009 objects which were not previously identified and which we list
as being 40\% to $>99$\% likely to be a QSO. As a byproduct of the
construction of this catalogue, we are able to list comprehensive
ROSAT field shifts as determined by our whole-sky likelihood
algorithm, and also plate-by-plate photometric recalibration of the
complete APM and USNO-A2.0 optical catalogues, significantly improving
accuracy for objects of $>15$ mag. The catalogue is available wholly and
in subsets at http://quasars.org/qorg-data.htm .} \maketitle

\section{Introduction}
In recent years a number of good-resolution radio and X-ray surveys
have been completed and the full data published. One major goal of
such surveys is that the radio/X-ray detections should be associated
with optical objects to further their classification and to find new
examples of emission phenomena. Previous such efforts generally treat
just one radio/X-ray survey per paper, and use matching criteria
particular to that paper; see notably APM Optical Counterparts to
FIRST Radio Sources (MWHB: McMahon et al. 2002) and the Hamburg/RASS
Catalogue of Optical Identifications (HRC: Bade et al. 1998) which has
multiple optical identifications per X-ray detection. It is desirable
for there to be a single unified catalogue which combines and overlays
all these good-resolution radio/X-ray surveys onto the optical
background using a uniform optimized matching algorithm. This paper
presents such a catalogue: the `Quasars.org' all-sky optical catalogue
of radio/X-ray sources, obtainable from
http://quasars.org/qorg-data.htm. The name refers to the website used
as a repository during this catalogue's development. We refer to our
catalogue as `QORG' throughout the rest of this paper.

The task of combining all these data was a complicated one, and our
general approach was to start with no preconceptions but to let the
data be our guide in evolving the best techniques. Iteration was
commonly used to find stable results for data merging and calibration
tasks. Extensive testing against well-understood control data allowed
us to develop heuristic solutions for {\it ROSAT} field shifting and
double radio lobe identification. We developed a whole-sky based
method of calculating likelihood-of-association which causally ties
optical objects to radio/X-ray sources. These likelihoods are written
into our catalogue as percentage odds that each associated optical
object is in turn a QSO, galaxy, star, or erroneous radio/X-ray
association. Objects presented are APM/USNO-A optical objects
calculated with $\ge 40$\% confidence to be associated with
radio/X-ray detections, or which are identified as known QSOs, AGN or
BL Lacs; the 40\% threshold is an arbitrary choice, but ensures that
the catalogue contains only interesting or potentially interesting
objects. These optical objects total 501,761 in all, including 119,816
objects bearing identifications from the literature and 86,009 objects
not hitherto identified which we list as being 40\% to $>99$\% likely
to be a QSO.

This paper is divided into sections as follows: (2) an account of all
the source catalogues used in this compilation; (3) a brief summary of
our primary likelihood algorithm, {\it ROSAT} field shifts, and
technique used to identify double radio lobes; (4) a description of
our main catalogue. The electronic paper includes an appendix
detailing, at some length, our methods and the issues encountered
during the construction of the catalogue. Its sections are: (1) issues
in the construction and recalibration of the merged optical catalogue
used for the background, and its attributes; (2) description of the
likelihood calculations used to causally associate optical objects
with radio/X-ray sources; (3) issues in overlaying the X-ray
detections onto the optical background, notably the field shifts
required; (4) issues in overlaying the radio detections onto the
optical background and identifying double radio lobes; (5) issues in
matching identification catalogues to the optical background; (6)
attributes and analysis of the resulting Quasars.org catalogue.

\section{The Source Catalogues}

Source catalogues included are categorized as optical, radio, X-ray, or
identification catalogues.

\subsection{Optical Surveys}
The whole-sky optical background represents by far the largest data
pool to be incorporated, although only those optical objects which are
associated with radio/X-ray detections, or are known quasars, are
included in the final QORG catalogue. This project commenced in
1999 and we used the optical data available at that time to compile
our own in-house whole-sky optical catalogue. Our main source was the
complete set of the Cambridge Automatic Plate Measuring machine (APM:
McMahon and Irwin 1992) scans of 1906 plates on the North and South
Galactic caps, consisting of 896 first-epoch National
Geographic-Palomar Observatory Sky Survey (POSS-I) $E$ and $O$ plates
centred on equatorial declinations $0\degr$ to $+90\degr$, and 1010
second-epoch UK Schmidt Telescope sky survey (UKST) ESO-R and SRC-J
plates centred on declinations $-85\degr$ to $0\degr$; these yielded
about 270,000,000 sources in one or more colours. We also include the
United States Naval Observatory whole-sky catalogues (USNO-A) which
used the Precision Measuring Machine (PMM) to read sources from the
POSS-I and UKST plates. The USNO-A catalogues are not as deep as the
APM so are treated as supplementary data, but only USNO-A covers the
Galactic plane area. The earlier USNO-A1.0 (Monet et al. 1998) lists
488,006,860 sources in both red and blue, with POSS-I plates used for
field centres down to declination $-30\degr$, and UKST plates below
that. USNO-A2.0 (Monet et al. 1998) lists 526,280,881 sources in both
red and blue; the additional sources were a result of a re-reduction
of the PMM scans and switching from POSS-I plates to the deeper UKST
plates for field centres with declinations of $-20\degr$ to $-30\degr$.

\subsection{Radio Surveys}
The largest radio survey is the NRAO VLA Sky Survey (NVSS:
Condon et al. 1998) catalogue 40 (2002), which is a 1.4-GHz all-sky survey
down to a declination of $-40\degr$, with a source detection threshold of
2.5 mJy and positional accuracy ranging from $<1$ arcsec for the strongest
sources to 7 arcsec at the faint limit.  A second radio survey is the
Faint Images of the Radio Sky at Twenty-cm survey (FIRST: White et al.
1997) which has recently (April 2003) been completed; this is a 1.4-GHz
survey of 9033 square degrees of primarily the north Galactic cap, with a
source detection threshold of 1 mJy and a positional accuracy within 1
arcsec.  The FIRST survey overlaps the NVSS in its surveyed area but is
deeper and has better resolution.  The part of the sky not covered by the
NVSS is currently being surveyed at 843 MHz by the Sydney University
Molonglo Sky Survey (SUMSS: Mauch et al. 2003, Oct 27 2003 release) to a
comparable depth and resolution; this survey is at this time about 70 per
cent complete so some of the sky below declination $-40\degr$ is as yet
without radio coverage to this resolution, but the total sky coverage
of these three radio surveys exceeds 95\%.

\subsection{X-ray surveys}

The best-resolution X-ray surveys up to the end of the last decade all
originate from {\it ROSAT} (ROentgen SATellite), which was operational
from 1990 to 1999; its extragalactic and Galactic surveys are
available in 4 primary catalogues. The {\it ROSAT} All-Sky Survey
(RASS / revision 1RXS) is derived from the all-sky survey performed
during the first half year of the {\it ROSAT} mission in 1990/91, and
is available as two separate sub-catalogues: the Bright Source
Catalogue (RASS-BSC: Voges et al. 1999a) containing 18,806 sources, and
the Faint Source Catalogue (RASS-FSC: Voges et al. 2000) containing
105,924 sources. The RASS has a sky coverage of 92\%, with a
nominal positional accuracy of 30 arcsec. Secondly, the {\it ROSAT}
Source Catalogue of Pointed Observations with the High Resolution
Imager (HRI / 1RXH: Voges \etal\ 1999b) final release 1.3.0 (2001) has
131,902 sources from 5393 sequences representing a sky coverage of
1.94\% with nominal positional accuracy of 5 arcsec. Third is
the Second {\it ROSAT} Source Catalogue of Pointed Observations with
the Position Sensitive Proportional Counter (PSPC / 2RXP: Voges \etal\
1999b) final release 2.1.0 (2001) with 116,259 sources from 5182
sequences, representing a sky coverage of 17.3\% with a nominal
positional accuracy of 25 arcsec. We include with this the
supplementary PSPC with Boron Filter catalogue (PSPCF: same
attributions as PSPC) release 2.0.0 (2001), with 2526 sources from 258
sequences representing a sky coverage of 0.15\%. Last is the
WGA Catalogue of {\it ROSAT} Point Sources (WGA: White, Giommi \&
Angelini 1994) final release (August 2000) with 115,962 sources from
4160 sequences, which covers the same observational data as 2RXP but
was originally released earlier and uses different data reduction
algorithms. We use the WGA catalogue in recognition of the role it has
played in research; it does include a few early sequences absent from
the PSPC catalogue.

\subsection{Identification catalogues}

The fullest description of any radio/X-ray emitting object in the
QORG catalogue is given when it is possible to identify it as a
known QSO, AGN, BL Lac, galaxy or star. The following are the source
catalogues for these types of objects which are used in the present
task; web sites describing many of these are listed in the online data
for the catalogue (http://quasars.org/ReadMe.txt)

The primary catalogue used for identification of QSOs, AGN and BL Lacs
is the Catalogue of Quasars and Active Nuclei, 11th edition (Veron:
V\'eron-Cetty \& V\'eron 2003) which identifies 64,866 such objects,
and uses an absolute-magnitude threshold to differentiate a QSO
classification from an AGN classification, to which we adhere. We have
added supplementary positional and name information from the large
recent releases of the Sloan Digital Sky Survey (SDSS: Abazajian
\etal\ 2003) and the 2dF QSO Redshift Survey (2QZ: Croom et al. 2003).
We have also added 52 extra QSO identifications from the NASA/IPAC
Extragalactic Database (NED) as those were found to have radio/X-ray
associations, and 11 extra QSOs from the SDSS quasar catalog 2nd
edition (Schneider et al, 2003) which made a supplementary release
based on re-inspection of the SDSS spectra too late for inclusion in
the Veron catalogue. However, we make use of only those objects for
which we have an optical counterpart; in total this gives 48,285 QSOs,
14,633 AGN and 841 BL Lacs.

A measured redshift is required for identification as a QSO, but
galaxies can reasonably be identified by visual morphology, although
spectroscopy remains decisive. The primary catalogue used for
identification of galaxies is the Principal Galaxy Catalogue (PGC)
which is extracted from the Lyon-Meudon Extragalactic Database (LEDA:
Paturel, Bottinelli, Gouguenheim 1995); our copy from September 2000
(courtesy of G. Paturel) contains 1,088,795 galaxies. We also use five
redshift surveys which make galaxy identifications over a large sky
area: the SDSS, the CfA Redshift Catalogue (CFA: Huchra et al. 1999,
April 2003 edition), the IRAS PSCz Redshift Survey (PSCz: Saunders et
al. 2000), the 2dF Galaxy Redshift Survey (2dFGRS: Colless et al.
2001) and the 6dF Galaxy Redshift Survey Early Data Release (6dFGS:
Wakamatsu et al. 2002). Some extra identifications are sourced from
the catalogue of Arcsecond Positions of UGC Galaxies (Cotton \& Condon
1999), the 2QZ, the online 3CRR catalogue at
http://www.3crr.dyndns.org/ (3CRR: Laing, Riley \& Longair 1983), the
Updated Zwicky Catalog (Zwicky: Falco et al. 1999) and the Redshift-
Distance Survey of Nearby Early-Type Galaxies (ENEAR: Wegner et al.
2003). To summarize, for galaxies not classified as AGN, we utilize
only those for which we have an optical object associated with a
radio/X-ray detection; these total 49,743 galaxies. Note that some
large galaxies known to be radio/X-ray emitters are missing from our
catalogue because of astrometric mismatches between the available
isophotally-bounded optical signatures and the radio/X-ray source
locations.

The remaining possibility is that objects are identified with stars.  This
has been somewhat problematic, in that until recently stellar
identifications were not often compiled, as they represented the detritus
of QSO or galaxy surveys.  Since radio/X-ray emitting objects are rarely
stars, if such an object displayed a star-like spectrum it may have served
only to keep it classified as an `unknown' object.  Large star catalogues
such as Tycho (Hog et al. 2000) are actually just point source catalogues
which do not make genuine stellar identification, and historic star
catalogues are too astrometrically imprecise for unambiguous computerised
matching, which we find to require astrometric precision of 15 arcsec or
better.  Recently, however, catalogues of stars of specific types such as
white dwarfs have been released to the required astrometric precision, and
large surveys like SDSS and 2dFGRS have published their star
identifications; thus in the last few years the availability of suitable
stellar data has greatly improved.  We have used the following star
catalogues for stellar identification: the Atlas of Cataclysmic Variables
(CV: Downes et al. 2001), Spectroscopically Identified White Dwarfs (WD:
McCook \& Sion 1999), the General Catalogue of Variable Stars (Vol 1) with
Improved Coordinates (GCVS: Samus et al. 2002), the revised New Luyten Two-
Tenths catalogue of high proper-motion stars (NLTT: Salim \& Gould 2003),
stars from the Large Bright Quasar Survey (LBQS: Hewett et al. 1995)
received courtesy of Paul Hewett, stars from the Las Campanas Redshift
Survey (Shectman et al. 1996), and star identifications from the galaxy
and QSO surveys listed above.  We have also included the Tycho survey, as
its objects are bright and very likely to be stars, and the Henry Draper
Extension Charts (HDx: Nesterov et al. 1995) even though their stars are
not confirmed spectroscopically.  We have obtained names of bright stars
from the Bright Star Catalogue, 5th Revised Ed. (Yale: Hoffleit \& Warren
1991) and the Common Name Cross Index (Smith W.B. 1996).  In the end we
utilize only those stars for which we have an optical object associated
with a radio/X-ray detection; these total 6314 stars.

\section{All-Sky Based Likelihood Calculations and Matching Techniques}

We give here a brief summary of the methods we used to relate optical
objects to radio/X-ray sources, and to identify double radio lobes. An
appendix that gives full details of our methods, together with
supporting tabulated data, can be found in the electronic version of
this paper.

Our primary algorithm to calculate the likelihood of association
between optical and radio/X-ray sources is based on
identifying classes of optical objects which tend to be
astrometrically co-positioned with radio/X-ray sources, and assessing
the significance of the relationship by comparison with whole-sky
background averages. For example, if a class of optical object is
found near NVSS sources at twice the areal density that it has on
average in the background, then we say that the chance of association
of those objects near the NVSS sources is 50\%, as we expect half of
the apparent associations to be chance superpositions of background
objects. We define these optical object classes using four parameters:
astrometric offset from the radio/X-ray source, photometric $(B - R)$
colour, APM psf classification, and local sky object density, binning
these to provide large populations in each class and so minimize
small-number fluctuations.

To improve the uniformity of our optical object classes we found it
necessary to recalibrate the source data. The APM plate depths were
photometrically recalibrated by matching stars on overlapping plate
margins; this was done separately for red and blue plates. USNO-A
photometry, which usually shows large zero-point offsets, was
recalibrated into the APM standard using matched stars. These
photometric recalibrations improve our $(B - R)$ colour data. The {\it
ROSAT} source positions were recalibrated by using our likelihood
algorithm to provide an optimal astrometric solution for each
sequence; these typically involved shifts of 1-10 arcsec on the sky.
These astrometric recalibrations improve our accuracy in gauging
positional offset between individual optical objects and X-ray
sources. As our recalibrations are potentially useful for others, we
provide them on-line: the APM/USNO-A2.0 recalibration is listed
plate-by-plate at
http://quasars.org/docs/QORG-APM-USNO-calibration.txt, and the {\it
ROSAT} field shifts are listed at
http://quasars.org/docs/HRI-fields.txt for the HRI catalogue, and
similarly for the RASS, PSPC and WGA input catalogues.

\begin{table}
\caption{Radio/X-ray Associations presented in the QORG catalogue.}
\begin{tabular}{lrrr}
\hline
Source catalogue& No. astrometrically & No. core &  No. double\\
                &  unique sources    &detections   &  lobes\\
&&                                     in QORG      &in QORG\\
\hline
   FIRST     &       781667     &     155132     &  11512\\
   NVSS     &       1810664     &     242851     &  8323\\
   SUMSS     &       165531     &      31156     &  1663\\
   HRI     &         56398     &      12733\\
   RASS     &        124730     &      30521\\
   PSPC     &        102005     &      29472\\
   WGA     &         88578     &      18712\\
\hline
\end{tabular}
\end{table}

As our aim was to derive maximum value from the source catalogues, we
have also endeavoured to identify double radio lobes from the radio
data.  As QORG is an optical catalogue, we are interested only in
those double lobes for which we have an optical centroid.  We used a
heuristic algorithm to identify these lobes, consisting of firstly
enumerating the likely lobe population inherent within the radio data,
then using a number of distinct rules to estimate the likelihood of a
given radio-optical-radio configuration being a member of that lobe
population.  The details are given in the appendix.  Table 1 summarizes the numbers of associations presented in QORG from each source catalogue.

\section{The Optical Catalogue of Radio/X-ray Sources}

The catalogue is available from the catalogue home page at
http://quasars.org/qorg-data.htm, and is written as one line per
optical object. The catalogue presents unique `best' associations, so
optical objects and radio/X-ray sources are not duplicated across
lines; this keeps the presentation simple and plain. The full
catalogue is in the `Master.txt' file (21Mb zipped) which provides
particulars of all 501,761 objects including data contributing to the
likelihood calculations and double lobe declarations. A condensed
version, `Free-Lunch.txt', is also provided; this displays no more
than 2 associations per object and omits supporting data. Also
available are two subsets, `Known-Objects.txt', which displays only
the 119,816 objects from our catalogue which are identified from the
literature, and `Quasar-Candidates.txt' which displays the 86,009
objects from our catalogue not hitherto identified which we list as
being 40\% to $>99$\% likely to be a QSO.

\begin{table*}
\caption{Sample lines from the QORG catalogue (`Free-Lunch' variant)}
\begin{tabular}{llllllrrrrllr}
\hline
J2000 location&type&$R$, $B$ (mag)&ct&psf&name& \multicolumn{4}{c}{type percentages}&$z$&radio/X-ray source 1&flux\\
(1)&(2)&(3) \ \ (4)&(5)&(6)&(7)&(8)&(9)&(10)&(11)&(12)&(13)&(14)\\
\hline    
040904.9-364744&GR &14.3 14.2 & &2 2&PGC 632512&0&98&0&2&&NVSS J040904.8-364745&113\\           040905.0-053236&RX &19.1 20.3 & &1 1& &12&74& 0&14& &NVSS J040904.6-053234& 4\\
040905.2-283859&R  &19.7 20.6 & &1 2& &21&56& 3&20& &NVSS J040905.3-283859& 4\\
040905.3+153056&R  &16.9 21.2 &p&n -& & 2&80& 2&16& &NVSS J040905.2+153051& 3\\
040905.4-092350&R  &17.2 19.4 &p&2 1& & 2&89& 0& 9& &NVSS J040905.4-092353&16\\
040905.8-123849&QR &18.0 18.4 &p&- -&PKS0406-127&97& 1& 0& 2&1.563&NVSS J040905.7-123847&450\\
040906.2-651733&R  &15.0 15.1 & &- -& &63&20& 3&14& &SUMSJ040905.3-651729&27\\
040906.2-041022&A  &18.5 19.9 &p&1 1&SDSSJ04-041& & & & &0.133\\
040906.3-760006&R  &13.0 13.6 & &- -& & 2&46&15&37& &SUMSJ040906.3-760006& 6\\
040906.5-051054&Q  &19.7 20.3 & &- -&SDSSJ04-051& & & & &1.556\\
040906.6-760534&R  &18.6 19.8 & &1 1& & 3&63& 0&34& &SUMSJ040906.7-760532& 6\\
040906.6+122356&X  &20.2(20.0)&p&2 x& & 0&57& 3&40& &2RXP J040906.9+122353& 6\\
040906.6+290944&SX &10.6 \ \ 0&p&n n&HD281690& 0& 6&64&30& &1RXS J040906.6+290943&92\\
040906.7-504531&R  &18.7 21.6 & &2 1& & 2&92& 0& 6& &SUMSJ040906.5-504528&18\\
040906.7-175710&QRX&19.1 20.6 & &- -&PKS 0406-18&64& 6& 4&26&0.722&NVSS J040906.6-175709&999\\
040906.8-681946&2  &11.7 11.5 & &- -& & 2&19&51&28& &SUMSJ040900.6-682023&36\\
040906.8-011844&R  &19.0 21.3 &p&1 -& &12&67& 0&21& &NVSS J040906.7-011845& 6\\
040907.3-043235&Q  &19.1 19.8 &p&- -&SDSSJ04-043& & & & &0.802\\
040907.6-304915&R  &20.6(22.5)& &- x& & 4&47& 9&40& &NVSS J040907.7-304916& 2\\
040908.0-695738&X  &18.8 21.5 & &n n& &18&56& 4&22& &1RXS J040907.9-695735&71\\
\hline
\end{tabular}
\end{table*}

\begin{figure*}
\epsfxsize 17.5cm
\epsfbox{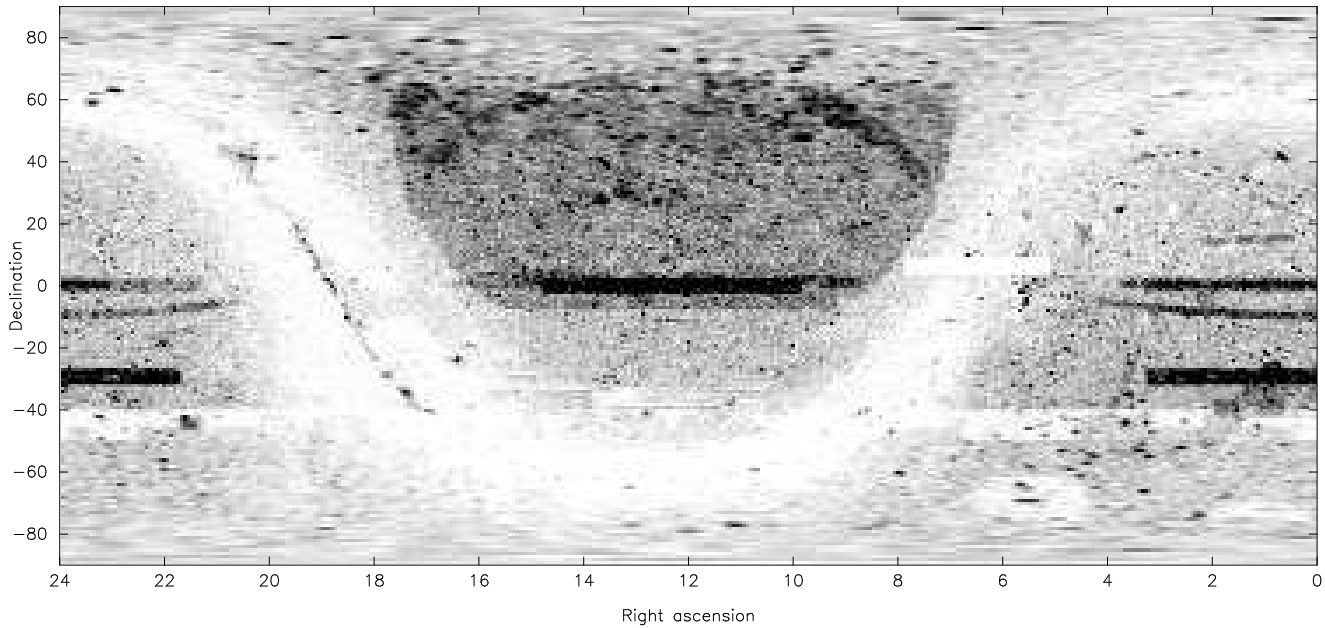}
\caption{A whole-sky optical density map of all 501,761 objects
presented in the QORG catalogue. See the text and the Appendix for
details. The pixels at high declinations have been stretched so that a
given shade of grey represents the same density of objects per unit
solid angle.}
\label{qorg}
\end{figure*}

Table 2 displays some sample lines of the QORG catalogue, using the
Free-Lunch version (which is easily tabulated while showing the
salient points of the similarly-structured main Master catalogue).
`ReadMe' files are provided on-line which give full file layouts,
field definitions and supporting information for all catalogues; we
only give an overview here. Column 1 displays the optical coordinates
(epoch J2000) which doubles as the IAU-recommended name of the object,
e.g., QORG J040904.9-364744. Column 2 summarizes any associations
with, and identification of, the optical object: R=radio source,
X=X-ray source, 2=double lobe declaration, Q=known quasar, A=AGN,
G=galaxy, S=star, B=BL Lac object. Columns 3 and 4 give the red and
blue magnitudes respectively, and column 5 states if those magnitudes
are POSS-I (='p') or UKST photometry, plus flagging any nominal
variability or proper motion. Column 6 gives the point spread function
(psf) classification of the two optical observations, taken largely
from the APM: '-'=stellar, '1'=fuzzy, '2'=extended, 'n'=no psf and
'x'=object not seen in this colour. Column 7 gives the name of the
object, where it is identified from the literature (abbreviated here
for space reasons). Columns 8-11 give the calculated probability that
the radio/X-ray associated object is turn a QSO, galaxy, star, or
erroneous radio/X-ray association; this is discussed further in the
next paragraph. Column 12 gives the redshift, if known. Column 13
gives the radio/X-ray source name for a declared association, and
column 14 gives the flux in mJy for a radio association, or the count
rate in counts/hour for a {\it ROSAT} X-ray association. A few of
these objects are, in the Free-Lunch catalogue, listed also with a
second radio/X-ray association which here is not shown for space
reasons. The Master catalogue, which we expect will be of most general
interest, lists up to six associations for each optical object,
together with particulars of any double radio lobe found for it,
supporting information which enables reconstitution of the likelihood
calculation for that object, and references to the source catalogues
for identified objects. Fig.\ \ref{qorg} is a whole-sky optical
density map of all 501,761 objects presented in the catalogue.

In the catalogue we display, for each radio/X-ray associated optical
object, the calculated probabilities that it is a QSO (including BL
Lacs), galaxy or star. We accumulated the data for these computations
from the identified optical objects in our catalogue, augmenting the
`star' pool with all unidentified optical objects which are 11th mag
or brighter. We placed objects classified as AGN into the QSO bin if
they had a stellar PSF in both colours, or where both colours were
fainter than 18.5 magnitude for USNO-A objects without PSF (there were
only 38 of these), and otherwise into the galaxy bin. Thus our
starting pool of known objects with radio/X-ray associations was 8628
QSOs, 52422 galaxies and 7078 stars. In separate exercises for the
radio and X-ray associations, we binned the associations by four
categories: radio/X-ray-to-optical astrometric offset (4 bins), $B -
R$ colour (16 bins), stellar APM PSF classification (4 bins), and
radio/X-ray-to-optical flux ratio (8 logarithmic bins); an additional
exercise omitting the PSF binning was done to cater for USNO-A sourced
objects which have no PSF data. The numbers of QSOs, galaxies and
stars are totalled within each cross-categorized bin; their ratios
will yield the relative likelihoods of each identification for that
bin. At least 20 objects are required for each bin to be usable; if
this was not the case, the bins were amalgamated until the 20 objects
are attained. However, this process yielded different results
depending on which categories were amalgamated first; we accommodate
this by amalgamating by eight primary sequences and taking the average
of the results. We ended up with ratios for each bin, of the form 53\%
QSOs, 36\% galaxies, 11\% stars. We then assigned those percentage
likelihoods to all radio/X-ray associated objects which belonged in
that bin, including the identified ones for comparison by the user
(objects associated with both radio and X-ray have their two results
combined), but for each individual object we also decrease those
percentages by the calculated chance that that object's radio/X-ray
association is false. This percentage chance of false association is
also listed, and the four percentages together add to 100\%; we round
the percentages to the nearest whole per cent, so a listed figure of
100\% is just a rounding rather than a statement of total confidence.
Objects thus given high QSO probability scores will be of the most
interest to researchers in the field; we enumerate 86,009 such objects in our
catalogue not hitherto identified which we list as being 40\% to
$>99$\% likely to be a QSO.  

The appendix, available in the electronic version of this article,
gives full details of all our methods along with supporting tabulated
data. The QORG catalogue and supporting data and ReadMe files can be
accessed from the catalogue home page at
http://quasars.org/qorg-data.htm .

\section{Summary}

This paper presents the QORG All-Sky Optical Catalogue of Radio/X-ray
Objects, which is intended to be a grand compilation of the
large-scale surveys of the radio and X-ray sky as they existed before
the beginning of {\it XMM} and {\it Chandra} operations. It uses the
completed {\it ROSAT}, NVSS and FIRST catalogues and the SUMSS
catalogue at 70\% completion. It provides optical associations
for these together with comprehensive identifications of known objects
with the intention of presenting an informative map to help formulate
and support pointed investigations.

\section*{Acknowledgements}

Sincerest thanks to Mike Irwin at Cambridge for providing documentation on
how to read APM files and for advice on calibration, to Dave Monet at the
USNO for generously providing copies of both versions of the USNO-A
catalog, and to Ray Stathakis at the AAO for coming to the rescue with a
large part of the APM UKST data which had hitherto been unobtainable, plus
copies of many AAO routines.  Thanks also to Rick White for clarifying
issues relating to FIRST astrometry and to Brian Skiff for discussions on
USNO-A photometry.  And great thanks to the people at arXiv.org who keep
science accessible to the wider public, without which this project would
have been much harder.

The National Radio Astronomy Observatory is a facility of the National
Science Foundation operated under cooperative agreement by Associated
Universities, Inc. The NASA/IPAC Extragalactic Database (NED:
nedwww.ipac.caltech.edu) is operated by the Jet Propulsion Laboratory,
California Institute of Technology, under contract with the National
Aeronautics and Space Administration.

MJH thanks the Royal Society for support.

\clearpage
\appendix
\section{Details of the catalogue construction}
\subsection{The Optical Catalogue used in QORG}

\begin{figure*}
\epsfxsize 17.5cm
\epsfbox{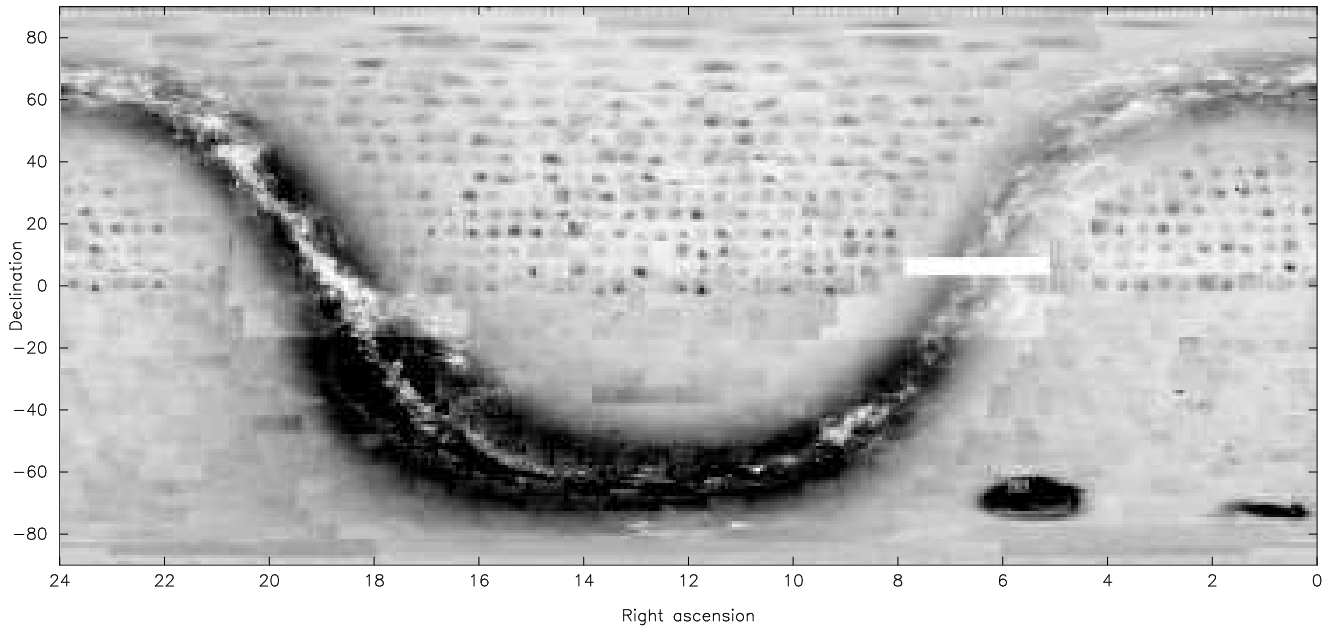}
\caption{A whole-sky optical density map of the sources in the optical
  catalogue. The missing sky coverage (white strip at 
centre-right) is due to corrupt USNO-A
  data; see the text for details.}
\label{skydens}
\end{figure*}

The APM and USNO-A catalogues have been combined into a whole-sky
670,925,779-object photometrically recalibrated catalogue. This was
done to provide an efficient and uniform optical background against
which to perform all other tasks. It was decided at the outset to
store astrometric positions to a precision of 1 arcsec only, as early
matching across APM plates showed typical discrepancies on the plate
margins of up to 2 arcsec from the mean, and we had no desire for
precision to exceed accuracy. The USNO-A catalogues have nominal
astrometric precision of 0.5 arcsec, but the APM astrometry was
selected where available because it is photometrically deeper than
USNO-A, and so should be used to ensure the best local astrometric
consistency of the merged data. Similarly, it was decided to store
photometry to a precision of 0.1 mag only, as early analysis across
APM plates showed 20\% of matching objects to have photometric
scatter greater than 0.3 mag, thus providing a sense of its accuracy.
Use of this modest precision standard enables our final optical
catalogue to be stored at just 7 bytes per object, converted to 11
bytes per object in our work files, which allows speedy processing for
whole-sky tasks. The density of objects on the sky in the resulting
catalogue is plotted in Fig.\ \ref{skydens}.

The APM and USNO-A present their data differently and, in a sense,
complement each other.  The APM classifies the point-spread function (PSF)
of each object as stellar, non-stellar (i.e. galaxy), merged, or non-
morphological, and seeks to display galaxy sizes, shapes, and position
angles by using ellipses to model isophotally-bounded areas.  The downside
of this is that close point-sources are often collected by the APM into a
`merged' object indistinguishable from a galaxy.  The USNO-A is oriented
to displaying stars so has no PSF classification and just describes point-
positions and magnitudes, but this means no distinction is made between
stars and galaxies.  By merging these two catalogues together, one gets
both kinds of information, and sometimes a bit extra.  APM `merged'
objects are often resolved by the USNO-A into constituent point sources.
Often photometry of different sections of a galaxy becomes available.  And
where an APM ellipse has a single USNO-A point source positioned at one
end of the ellipse with no other USNO-A object present, the properties of
an object at the other end can be calculated; comparison with Digitized
Sky Survey (DSS) images show that the calculated object is correct to
within a few arcsec in position and 1-2 mag photometrically.  Such
objects have been included in our optical catalogue and are flagged as
`inferred objects'.  Any APM `merged' object that we have resolved into
constituent point sources is dropped while the resolved sources are
included in our optical catalogue.

Some issues encountered in reading the APM data were:
\begin{enumerate}
\item Some APM plates were missing their calibration parameters, so
default values were supplied which were later adjusted in the
subsequent whole-sky calibration exercise. \item About 10 of our
1997-dated POSSI-based files were missing J2000 coefficients in the
headers. This was remedied by mapping individual objects from the
B1950 positions using the transformation matrix from Murray (1989)
which was found to yield J2000 positions accurate to within the
required arcsec precision. \item Overly-flattened ellipses were found
to be spurious signals. A threshold was established to remove such
objects. Also, the APM has a photometric classification for
static-like (non-morphological) signals; it was found that objects
having only this classification were usually false positives and so
were removed. We felt that any true objects thus lost would generally
be restored with the subsequent addition of the USNO-A data. \item
Many point sources are seen in only one colour as the counterpart of
the other colour is fainter than the plate depth. Sometimes, however,
a point source in one colour has its counterpart of the other colour
concealed within a `merged' ellipse with an offset centroid, so
appearing to be entirely missing in that colour. We felt it important
to distinguish between such concealment and genuine absence, so in
such cases we have filled out the object data by adding the ellipse
photometry for the missing colour.\item About half of the POSS-I
plates contain spurious one-colour `objects' positioned preferentially
toward the plate centres; this is evident on the optical density chart
of Fig.\ \ref{skydens}. They are an artefact on the glass copies of
the POSS-I plates which originated from defects in the older 103aE and
O emulsions that were most strongly imaged in the central area during
the copying process. These are very faint but were detected by the
deep APM scans of those glass copies (M. Irwin, private
communication). In worst cases these can double the nominal population
of a POSS-I plate, but they have been found via pattern analysis to
have had no discernible effect on our efforts; we have probably
benefitted from our approach of matching optical objects to
radio/X-ray detections, which also confirms that the matched object is
likely to be real. See MWHB section 3.5, where they similarly find
that FIRST detections confirm matching APM `noise' objects as likely
to be real.
\item Large isophotal ellipses within large galaxies can be
astrometrically misaligned between red and blue plates, causing APM to
display neighbouring pairs of notional one-colour or mismatched-colour
`objects', one blue and the other red, both non-stellar. There was no
simple fix for this which would not introduce errors, so such data
within large galaxies originate from this artefact. \item To allow
easy reference from a lookup table, we chose to crop each APM plate to
the maximal simple rectangle of sky bounded by two longitudes and two
latitudes (J2000) - some care was needed in this to avoid loss of sky
coverage, i.e. each cropped plate must at least reach all its
neighbours. This task was made more delicate by the fact that the
original plates were arrayed by B1950 coordinates which are at a small
angle to our J2000 boundaries. \item An APM plate solution designed to
correct astrometric plate distortion is available, but we chose to use
the raw APM astrometry due to the complex nature of the solution. In
this we feel justified by the findings of MWHB that the plate solution
actually increases offsets of faint objects near the plate corners. In
general our raw APM astrometry is correct to an error of 1 arcsec in
RA and DEC, with occasional errors of 2 arcsec in RA and DEC as
determined by comparison to FIRST astrometry; see MWHB for a full
discussion of these issues.
\end{enumerate}

Some issues encountered in using the USNO-A data were:
\begin{enumerate}
\item At the
POSS-I and UKST source-plate boundaries (within the USNO-A data) it
frequently occurs that an object is represented twice, being on both
sides of the boundary. Such duplicate objects within a 4-arcsec
separation were removed. \item Data for 17 northern-sky POSS-I plates
were found to be corrupted in both A1.0 and A2.0 catalogues, i.e.
basically empty of data there. The affected area is bounded roughly by
RA 5h-12h and DEC $3\degr$ -- $8.3\degr$. Half of this is covered by
the APM, leaving the area bounded by RA 5.6h-8.3h and DEC
$3\degr$--$8.3\degr$ (about 243.7 sq deg, 0.59\% of the sky)
without coverage in our optical catalogue. \item Similar corruption
occurs in 17 southern-sky plates in the A2.0 catalogue. Fortunately
the A1.0 catalogue has no problems here, so it was used to populate
this region of sky. Oddly, the affected USNO-A plates are those
numbered 537 -- 553 in each hemisphere. \item There are substantial
photometric zero- point offsets in the A2.0 catalogue; the listed
values are nearly a full magnitude too bright, except for red POSS-I $E$
data. The problem was remedied via calibration into APM-governed
magnitude ranges. The A1.0 catalogue is not thus affected and seems
well calibrated. \item Southern- sky POSS-I plates displayed a
systematic pattern of objects being 0.3 mag fainter at the south end
of each plate compared with objects at the north end. This presumably
results from the thicker sky cover at lower angles.
\end{enumerate}

Our optical catalogue was initially assembled one APM-based plate at a
time by adding in corresponding data from the USNO-A2.0 catalogue, as
well as USNO-A1.0 as needed. Objects were matched across input
catalogues to a separation of 3 arcsec in each of RA and DEC
regardless of photometry, while accommodating best fits for objects
multiply packed more closely together. Intra-plate photometric
calibration was done separately for red and blue by establishing the
median offset between the APM and USNO-A2.0 data, then adjusting the
USNO-A2.0 magnitudes by that amount to attain the APM standard; this
was done separately for USNO-A1.0 data where we used it. Our optical
catalogue retains only a single red and blue magnitude value for each
object, so the APM photometry was retained as the first choice in all
cases except when the only available POSS-I photometry was from
USNO-A, as POSS-I magnitudes are preferred. This is because (a) POSS-I
$E$ (red) and $O$ (blue) plates were photographed on the same night,
thus ensuring the colour magnitudes are comparable. By contrast, UKST
$R$ (red) and $Bj$ (blue) plates are often obtained e.g. 10 years
apart, so variability can spoil the colour comparison. (b) POSS-I $O$
is centred on violet, 4050\AA, making a broader colour baseline with
the red 6400\AA\ (for both POSS-I $E$ and UKST $R$) than does UKST $Bj$
4850\AA. We have found, from 2227162 stellar objects on overlapping
equatorial POSS-I / UKST plates after calibration, that the median
value of $(Bj - R) / (O - E)$ was 0.65.

After assembly of 824 two-colour APM-based plates (i.e. all those
available in 1999, with two overlapping North pole plates treated as a
single plate), next came the task of whole-sky photometric calibration.
The APM photometry was recalibrated plate by plate by comparing magnitude
values of matched objects on cropped plate overlaps, rolled up into a
median offset for each two-plate combination.  The POSS-I plates were
calibrated together in one exercise, the UKST in another.  Objects used
were those of stellar PSF in both colours on both plates and with
positions that agreed to within 2 arcsec inclusive in both RA and DEC -
the closer criterion was used to ensure true matches.  Calibration was
done by adjusting all plate magnitudes by half of the indicated amounts
from overlapping areas, then repeating until near-stability was reached,
i.e. to where the absolute change per plate averaged less than 1/200th of
a magnitude.  This took 15 iterations to achieve for the POSS-I plates,
and 10 iterations for the UKST plates. The photometric scatter about the
median offsets is displayed in Table 1, astrometric scatter in Table 2.
The final magnitudes were rounded to 0.1 mag, as described above.

The calibrated magnitudes of objects from APM POSS-I plates were found to
vary from the nominal values mostly within a range of $\pm 0.4$ mag, but
discrepancies of up to a full magnitude were found.  The UKST plates were
more stable.  The calibrated APM POSS-I $E$ plates were found to have a zero-
point offset of 0.2 mag compared with the UKST; that is, the $E$ plates were
nominally on average 0.2 mag too bright.  After confirmation (Mike Irwin,
private communication), all POSS-I $E$ magnitudes were made 0.2 mag fainter.
The outcome of the full calibration shows that POSS-I plates are often
considerably deeper than the nominal magnitude limit.  An extreme example
is eo789 which calibrates as having a depth of $E=21.2$ and $O=22.7$, easily
deeper than the POSS-II coverage there, confirmed by examining DSS images.
Of course, other POSS-I plates can turn out quite shallow, e.g. eo774 with
a depth of $E=19.1$ and $O=20.2$.  One particularly notable result was that
the Large Magellanic Cloud plate f056 was calibrated into being over a
full magnitude brighter than APM nominal.  The 3823 overlapping stars
which yielded this result were carefully examined, and the offset was
found to be uniform with normal scatter.  The brighter LMC magnitudes are
included in our optical catalogue.

134 additional two-colour APM plates were obtained in March 2002, all but
one in the southern hemisphere, and these were added by reconstituting the
final catalogue in those places using the same processing rules.  These
new plates were calibrated to the QORG baseline by comparing
stellar objects on overlapping plate margins and simply adjusting by the
offset median. Our calibration is listed plate-by-plate at
http://quasars.org/docs/QORG-APM-USNO-calibration.txt, which also lists the
MWHB POSS-I E calibration of 148 of these APM plates using APS.  Table 3
summarizes this calibration of all 958 APM-based plates.

It remained to calibrate the large Galactic plane area, which is
covered only by the USNO-A. The APM-based plates showed the median
adjustments for USNO-A2.0 were to add +0.2 to POSS-I $E$ and +0.8 to
POSS-I $O$, and +0.9 to UKST R and +0.7 to UKST $Bj$; see the
aggregate summary in Table 3. These offsets were applied to all
USNO-A2.0-only areas, except that north of declination $+63\degr$ the
local APM-based plates indicated a POSS-I $O$ adjustment of just +0.3;
the half-magnitude difference indicates the limit of our ability to
bulk calibrate the USNO-A data in the absence of co-positioned APM
data. These Galactic plane adjustments completed the photometric
recalibration of our optical catalogue.

Preparatory to assembling our all-sky catalogue, we needed to
integrate the APM-based equator which is covered by both POSS-I and
UKST plates. We combined these by matching objects with positions that
agreed to within a separation of 3 arcsec inclusive in each of RA and
DEC. The UKST plates are generally deeper than POSS-I plates and so
have more objects; thus, we use UKST astrometry where available to
preserve local astrometric consistency and provide the most recent
position, but we use POSS-I photometry where available, although
two-colour UKST objects were chosen over one-colour POSS-I objects.
Therefore the result of combining these is an interwoven mix of POSS-I
and UKST objects and attributes, with a flag to indicate where the
blue magnitude is POSS-I $O$. In this way, 29 equatorial POSS-I plates
and 24 UKST plates were entirely written onto their counterparts and
so not further used. Similarly, the USNO-A1.0 has POSS-I coverage
between $-17\degr$ and $-33\degr$ which is covered in UKST by
USNO-A2.0, so the POSS-I data was overlaid onto the UKST background
and internally calibrated by adjusting both $E$ and $O$ by the median
$(R-E)$ offset for each two-plate combination; this method keeps
POSS-I $O$ and UKST $Bj$ photometrically distinct.

The remaining task was to combine all plates into continuous data
covering the sky. The recalibrated USNO-A was initially used as the
background, to be tiled over by the APM-based plates. Where plates
overlap, it is desirable to use the deepest plate; we therefore
ordered the plates from lowest plate depth to highest and tiled them
onto the background in that order. The deeper plates thus overwrite
the shallower ones. Merging was performed at the plate boundaries to
ensure no object was lost, as well as de-duplication to a separation
of 3 arcsec in each of RA and DEC. Post-assembly analysis revealed
some small `holes' in the sky coverage which were manually repopulated
from whichever APM plate had the data. As mentioned, the astrometric
precision of the final optical catalogue is to one arcsec only.
This allocates 1,296,000 R.A. units along the equator. These units
naturally compress toward the celestial poles. To ease processing, we
allocate only 432,000 R.A. units between declinations $60\degr$ and
$75\degr$, 259200 R.A. units between declinations $75\degr$ and $85\degr$,
and just 86400 R.A. units poleward of declination $85\degr$. These
roundings conform to the 1 arcsec astrometric precision for which we
are aiming.

The finished optical catalogue has 155,108,493 POSS-I sources and
112,827,180 UKST sources from the APM, and 192,176,786 POSS-I sources and
210,533,717 UKST sources from USNO-A.  These crisp photometric totals mask
the fact that many of these QORG optical objects are two-epoch hybrids
having POSS-I photometry and UKST astrometry.  There are in addition a
total of 279,603 inferred objects which appear only in this catalogue,
133,018 inferred from POSS-I data and 146,585 from UKST data.  As there is
no PSF information on inferred objects we treat them as non-APM except for
286 which are matched in the other colour to an unresolved off-centre APM
`merged' ellipse and so are treated as APM-type due to their nominal PSF.
All of these add up to 670,925,779 unique objects in the QORG
optical catalogue, which maps the sky north of $+3\degr$ in POSS-I, south of
$-33\degr$ in UKST, the Galactic plane north of $-17\degr$ in POSS-I, and the
remainder in a two-epoch mix of both.  A comprehensive
listing of individual cropped-plate sky boundaries, plate depths, and
counts of the objects categorized by PSF type can be found in the file
http://quasars.org/docs/QORG-plate-summary.txt .  Table 4 displays the
object totals for our optical catalogue where each of our two-colour
processed plates is allocated wholly by survey (POSS-I/UKST/BOTH) and
source catalogue (APM/USNO-A).  These finished processed plates share no
objects with their neighbours and can have irregular boundaries and
residues of objects from adjacent areas, e.g., the POSS-I plates can
contain some UKST objects where they border on UKST areas.  These
aggregate totals summarize the integrated optical catalogue that we have
used throughout this project.

Of particular note in Table 4 are the two-epoch objects.  Our optical data
retains no explicit two-epoch flag (except where the object is flagged as
variable or having proper motion), but since we retain the POSS-I
photometry for all such two-epoch matches, and $\sim 98$\% of POSS-I
objects in this sector have UKST counterparts, we can make the general
statement that all objects in this sector annotated as POSS-I are two-
epoch in our catalogue, and UKST objects are not, i.e., there was no good
POSS-I match for those UKST objects.  An exception is the equatorial
plates that were covered by APM in both POSS-I and UKST; here we find that
about 16\% of the flagged 2-epoch objects are in fact UKST from
overlapping APM SERC plates.  Additional two-epoch objects come from such
overlaps of our cropped APM-based plates, for which we evaluated only
objects that were stellar in both colours when calibrating our optical
catalogue; we retained only those two-epoch objects from the APM overlaps.
APM POSS-I plates are $6.5\degr$ on a side and positioned at $6\degr$
intervals, so two-epoch areas are small to begin with; after our cropping
and object selection we retained only 1.2\% of all objects as two-
epoch, as shown in Table 4.  The APM UKST plates are also $6.5\degr$
square but are positioned at just $5\degr$ intervals, which optimally
allows 70\% two-epoch coverage; however, the usefulness of two-
epoch UKST coverage is tempered by the UKST red and blue images being
taken at different epochs, so that variability and proper motion can be
jumbled and lost; after our cropping and object selection we retained only
8.7\% of all objects as two-epoch.  In total 10.7\% of our
optical catalogue objects are sourced from two epochs, comprising 18 per
cent ($\sim62200000/347418297$) of POSS-I objects and just 3\%
($\sim 10000000/323507482$) of UKST objects; the prevalence of POSS-I two-epoch
objects, again, is a consequence of our systematic retention of two-colour
POSS-I photometry wherever available.

A token effort was made to detect variability and proper motion across
epochs in our data prior to the final assembly of our optical
catalogue. Matched objects with post-calibration variability of over
1.0 mag (exclusive) in each colour have been flagged as variable,
although where both epochs were APM then the threshold is 0.5 mag
because of the uniformity of the calibrated APM photometry. We flag
3,702,933 such objects in our complete two-epoch zone between
declinations $+3\degr$ and $-33\degr$, comprising about 5.7\%
of all objects there. Testing of GCVS stars (for which there is no
published completeness) in our two- epoch zone shows we flag 283 out
of 851 GCVS stars there as variable for a 33\% identification
rate, which is a fair result given that many of these stars will have
been at equivalent points of their light curves in both epochs, or at
different points of their light curves for the discrete epochs of the
UKST-R and UKST-Bj plates, which would confuse the comparison to the
POSS-I data. In regard to proper motion, matched stellar objects with
post-astrometric-calibration positional shifts of 3-8 arcsec have been
flagged in our optical catalogue as displaying proper motion; these
total 871,705, comprising 1.2\% of all our two-epoch objects.
We have tested our results against those stars from the Tycho and NLTT
surveys which are listed with proper motions of $>0.08$ arcsec/year
which should show up as a 3-arcsec shift across the $\sim 30$-year
span of our two epochs. We test against Tycho stars in our complete
two-epoch zone (as with GCVS) and our optical catalogue flags 6753 out
of 15515 qualifying Tycho stars as moving, for a 43.5\%
identification rate, which seems low; however, these are bright stars,
many of which were astometrically inserted into the USNO-A instead of
using standard PMM reductions. The NLTT lists faint moving stars
perhaps more suited to comparison to our optical catalogue; it has
36,085 stars, being 90\% complete over 44\% of the sky.
Testing against the NLTT over the entire sky shows our optical
catalogue flags as moving 3402 out of 33,975 qualifying NLTT stars
that we find in our catalogue. As our whole- sky two-epoch
completeness is just 10.7\% this indicates a $\sim 93$\%
($3402/(33975\times.107)$) identification rate of NLTT stars as
moving. While at first glance this looks pretty good, further
inspection shows that the completeness of NLTT indicates that there
should be only about 91,000 such high proper-motion stars over the
whole sky, whereas we flag 871,705 such objects, so we have about ten
times too many. By comparison, Gould (2003) notes that the USNO-B
catalogue (Monet et al. 2003) flags one hundred times too many high
proper-motion stars compared with NLTT, but the USNO-B authors elected
to over-report as a method of designating high proper motion
candidates. Our goal was simply to accurately identify these objects,
so it seems that we have overreached somewhat. Our partial success in
flagging variable and proper motion objects shows that these flags
should be taken as indicative only, and needing confirmation in
individual cases.

\begin{table*}
\caption{Photometric scatter about the median offset for matched
objects on overlapping APM plates. All included objects have stellar
PSF in both colours on both plates.}
\begin{center}
\begin{tabular}{rrrrrrrrr}
\hline
&  \multicolumn{2}{c}{POSS-I $E$} & \multicolumn{2}{c}{POSS-I $O$}&\multicolumn{2}{c}{UKST $R$}  &\multicolumn{2}{c}{UKST $Bj$}\\
Magnitude&Number of&Cumulative&Number of&Cumulative&Number
of&Cumulative&Number of&Cumulative\\
difference&matches&percentage&matches&percentage&matches&percentage&matches&percentage\\
\hline
  0.0  &   256530   &   16.24   &  232634   &   14.73  &  1317056&
24.32  &   985544   &   18.20\\
  0.1  &   446552   &   44.52   &  416090   &   41.07  &  2039190&
61.98  &  1703200   &   49.65\\
  0.2  &   311352   &   64.23   &  309549   &   60.67  &  1061705&
81.58  &  1149693   &   70.88\\
  0.3  &   192744   &   76.43   &  204464   &   73.62  &   483864&
90.52  &   668240   &   83.22\\
  0.4  &   117074   &   83.85   &  129618   &   81.83  &   228729&
94.74  &   371877   &   90.09\\
  0.5  &    72618   &   88.45   &   84078   &   87.15  &   115546&
96.87  &   206804   &   93.91\\
  0.6  &    47723   &   91.47   &   56436   &   90.72  &    60490&
97.99  &   118176   &   96.09\\
  0.7  &    32845   &   93.55   &   39041   &   93.20  &    33789&
98.61  &    68602   &   97.35\\
  0.8  &    22946   &   95.00   &   27518   &   94.94  &    20956&
99.00  &    41267   &   98.12\\
  0.9  &    16581   &   96.05   &   19822   &   96.19  &    13467&
99.25  &    25972   &   98.60\\
  1.0  &    12418   &   96.84   &   13829   &   97.07  &     9301&
99.42  &    16849   &   98.91\\
  1.1  &     9589   &   97.44   &   10237   &   97.72  &     6590&
99.54  &    11844   &   99.13\\
  1.2  &     7298   &   97.90   &    7550   &   98.20  &     4833&
99.63  &     8866   &   99.29\\
  1.3  &     5730   &   98.27   &    5679   &   98.56  &     3716&
99.70  &     6827   &   99.42\\
  1.4  &     4699   &   98.56   &    4204   &   98.82  &     2781&
99.75  &     5554   &   99.52\\
  1.5  &     3754   &   98.80   &    3307   &   99.03  &     2235&
99.79  &     4383   &   99.60\\
  1.6  &     3014   &   98.99   &    2717   &   99.20  &     1794&
99.83  &     3570   &   99.67\\
  1.7  &     2509   &   99.15   &    2151   &   99.34  &     1436&
99.85  &     3037   &   99.72\\
  1.8  &     2083   &   99.28   &    1770   &   99.45  &     1209&
99.88  &     2532   &   99.77\\
  1.9  &     1788   &   99.40   &    1410   &   99.54  &     1010&
99.89  &     2134   &   99.81\\
  2.0+ &     9518   &  100.00   &    7261   &  100.00  &     5704&
100.00 &     10430  &   100.00\\
\hline
 Total &  1579365  &&             1579365  &&             5415401&&
5415401\\
\hline
\end{tabular}
\end{center}
\end{table*}

\begin{table*}
\caption{Astrometric scatter about the median offset for matched stellar
objects on overlapping APM plates.
          All included objects have stellar PSF in both colours on both
plates.
          Note: selection effect at 3 arcsec; multiply number of objects by 3 to
obtain true background approx.}
\begin{tabular}{rrrrrrrrr}
\hline
Scatter   &   \multicolumn{4}{c}{POSS-I} &  \multicolumn{4}{c}{UKST}\\
(arcsec) &  Number in Dec. & Percentage & Number in RA& Percentage &
 Number in Dec. & Percentage & Number in RA& Percentage\\
\hline
    0   &   848948  &   53.75  &    691329  &   43.77  &    3376931  &   62.36&
2977660 &    54.99\\
    1   &   663282  &   42.00  &    785769  &   49.75  &    1991025  &   36.77&
2297920 &    42.43\\
    2   &    65641  &    4.16  &    100762  &    6.38  &      47338  &    0.87&
139132  &    2.57\\
    3   &     1494 &     0.09   &     1505  &    0.10  &        107  &    0.00&
689     & 0.01\\
\hline
Total &  1579365  &  100.00  &   1579365 &   100.00  &    5415401 &   100.00
& 5415401  &  100.00\\
\hline
\end{tabular}
\end{table*}

\begin{table*}
\caption{Photometric calibration of the APM and USNO-A2.0 catalogue
summarized by plate depth adjustment.  The three right-hand columns
compare the calibration of 148 POSS-I E plates by MWHB and this paper.
Columns are as follows
 (1)  classification:  magnitude amount added to plate depth to obtain new
plate depth
 (2)  Number of POSS-I $E$ plates, from APM depth to QORG depth
 (3)  Number of POSS-I $O$ plates, from APM depth to QORG depth
 (4)  Number of UKST $R$ plates, from APM depth to QORG depth
 (5)  Number of UKST $Bj$ plates, from APM depth to QORG depth
 (6)  Number of POSS-I $E$ plates, from USNO-A2.0 depth to QORG depth
 (7)  Number of POSS-I $O$ plates, from USNO-A2.0 depth to QORG depth
 (8)  Number of UKST $R$ plates, from USNO-A2.0 depth to QORG depth
 (9)  Number of UKST $Bj$ plates, from USNO-A2.0 depth to QORG depth
(10)  148 POSS-I E plates, from APM depth to MWHB depth (rounded to 0.1
mag)
(11)  The same 148 plates, from APM depth to QORG depth (this is a
subset of column 2)
(12)  The same 148 plates, from MWHB depth to QORG depth.}
\begin{center}
\begin{tabular}{rrrrrrrrrrrr}
\hline
   (1)   &   (2)   &   (3)   &   (4)    &  (5)  &    (6)   &   (7)  &    (8)&
(9)   &  (10)  &   (11) &    (12)\\
\hline
$  -1.4$   &     .    &    .    &    .   &     1   &     .   &     .    &    .&
   &     .   &     .    &    .\\
  $-1.2$   &     .    &    1    &    .   &     .   &     .   &     .    &    .&
   &     .   &     .    &    .\\
  $-1.1$   &     .    &    1    &    .   &     .   &     .   &     .    &    .&
   &     .   &     .    &    .\\
  $-1.0$   &     1    &    1    &    1   &     2   &     .   &     .    &    .&
   &    .    &    .     &   .\\
  $-0.9$   &     1    &    .    &    .   &     .   &     .   &     .    &    .&
   &    .    &    .     &   .\\
  $-0.8$   &     .    &    .    &    .   &     .   &     .   &     .    &    .&
   &    1    &    .     &   .\\
  $-0.7$   &     1    &    2    &    1   &     4   &     .   &     .    &    .&
1   &    .    &    .     &   .\\
  $-0.6$   &     3    &    7    &    .   &     6   &     .   &     .    &    .&
   &    2    &    1     &   .\\
  $-0.5$   &     3    &    8    &    6   &    11   &     .   &     .    &    .&
   &     1   &     1    &    .\\
  $-0.4$   &     3    &   21    &   15   &    18   &     .   &     .    &    .&
   &     3   &     .    &    3\\
  $-0.3$   &     5    &   29    &   36   &    46   &     .   &     .    &    .&
   &     6   &     .    &    3\\
  $-0.2$   &    25    &   54    &   60   &    39   &     .   &     .    &    .&
   &     5   &     5    &   17\\
  $-0.1$   &    34    &   61    &   89   &    92   &     7   &     .    &    1&
   &    13   &    10    &   25\\
   0.0   &    48    &   73    &  123   &   105   &    49   &     .    &    .&
   &    17   &    19    &   29\\
   0.1   &    60    &   60    &   91   &    91   &   102   &     .    &    .&
1   &    11   &    17    &   28\\
   0.2   &    74    &   57    &   45   &    43   &    90   &     4    &    .&
   &    23   &    27    &   24\\
   0.3   &    68    &   47    &   22   &    25   &    84   &     2    &    .&
4   &    20   &    21    &   11\\
   0.4   &    50    &   16    &   10   &    13   &    57   &     4    &    7&
15  &     19  &     16   &     4\\
   0.5   &    40    &    4    &    7   &     5   &    34   &    26    &    3&
26  &     14   &    20   &     1\\
   0.6   &    19    &    4    &    2   &     7   &    12   &    59    &   22&
63  &      8   &     7   &     3\\
   0.7   &     6    &    .    &    1   &     1   &     2   &   110    &   42&
71  &      3   &     2   &     .\\
   0.8   &     4    &    1    &    1   &     1   &     1   &    90    &   46&
53  &      1   &     2   &     .\\
   0.9   &     2    &    1    &    .   &     .   &     .   &    62    &   65&
37  &      1   &     .   &     .\\
   1.0   &     .    &    .    &    .   &     .   &     .   &    39    &   42&
16  &      .   &     .   &     .\\
   1.1   &     .    &    .    &    .   &     .   &     .   &    22    &   25&
7   &     .    &    .    &    .\\
   1.2   &     1    &    .    &    .   &     .   &     .   &    15    &   23&
7   &     .    &    .    &    .\\
   1.3   &     .    &    .    &    .   &     .   &     .   &     2    &   13&
5   &     .    &    .    &    .\\
   1.4   &     .    &    .    &    .   &     .   &     .   &     2    &   11&
   &     .    &    .    &    .\\
   1.5   &     .    &    .    &    .   &     .   &     .   &     1    &    4&
   &     .    &    .    &    .\\
   1.6   &     .    &    .    &    .   &     .   &     .   &     .    &    1&
   &     .    &    .    &    .\\
   1.7   &     .    &    .    &    .   &     .   &     .   &     .    &    1&
   &     .    &    .    &    .\\
\hline
 Total   &   448  &    448  &    510  &    510  &    438  &    438   &
 306 &
306   &   148  &    148    &  148\\
\hline
\end{tabular}
\end{center}
\end{table*}

\begin{table*}
\caption{Counts of optical objects in the QORG optical catalogue,
subdivided by POSS-I/UKST surveyed areas (`BOTH' indicates 2-epoch areas)
and source APM/USNO-A catalogue.  Note that all 2-epoch objects from areas
surveyed by both POSS-I and UKST are POSS-I, as POSS-I photometry was
always retained for these.}
\begin{tabular}{llrrrrrrr}
\hline
        &  Source   &  No. of $R$+$B$  &  Area  &No. of optical   & No. of
        POSS-I &  No. of UKST  & No. of 2-epoch  &  2-epoch\\
Survey &  Catalogue &  plates &  (sq deg)  &   objects  &    objects&
objects &     objects  &percentage\\
\hline
 POSSI&  APM \& USNO-A &   448 &   13504.9  &  133053261 &   133046581&
6680   &   1579365   &    1.2\\
 POSSI&  USNO-A only  &   296  &   5799.1  &  149390371 &   149204938&
185433   &         0     &    0\\
 UKST &  APM \& USNO-A &   201  &   4534.7  &   62582083   &     20972&
62561111  &    5415401    &   8.7\\
 UKST &  USNO-A only   &  207   &  4857.8 &   170296427    &    45907&
170250520  &          0    &     0\\
 BOTH &  APM \& USNO-A &   309  &   7977.7  &   82968412  &   27932578&
55035834  &  $\sim 28000000$  &    33.7\\
 BOTH &  USNO-A only   &   76   &  4335.0  &   72635225   &  37167321&
35467904  &  $\sim 37000000$  &    51.0\\
\hline
 TOTAL   &  &            1537 &   41009.3  &  670925779  &  347418297 &
323507482 &   $\sim 72000000$ &      10.7\\
\hline
\end{tabular}
\end{table*}

\subsection{Calculation of the Likelihood of Association between Optical Objects and
Radio / X-ray Sources}

The distinguishing technique of the QORG catalogue is the uniform
algorithm by which likelihood of association between optical and radio/X-
ray sources is calculated.  The na\"\i ve approach to causal linking of these
would be to search for simple astrometric co-positionality, but problems
with that approach include the natural offsets in extended objects and
jets and lobes, the astrometric imprecision of the available data,
especially the X-ray data, and the differing significance of co-
positionality in dense star fields compared to sparse.  The FIRST Bright
Quasar Survey (FBQS: White et al. 2000) aligned radio and optical
astrometry to a precision of 0.1 arcsec and found that co-positionality
was a sufficient sole criterion for association only out to a 1.2 arcsec
separation in sky areas away from the Galactic plane.  The present work
treats positional separation only in increments of 1 arcsec, and uses this
with additional criteria to quantify likelihood of association.  As an
example, given two equivalent nearby optical candidates for association
with a radio/X-ray source, if one of them has R = B and the other has R =
B - 2.5, we would consider the former to be the far more likely candidate
as it has QSO-like colours, while the other is likely to be a coincident
star.  But to weigh this distinction accurately requires quantitative
assessment of the likelihoods to be assigned to different optical colour
bins.  In total we use three observational parameters to assess the
likelihood of association between radio/X-ray sources and optical
candidates: astrometric offset, B - R colour, and APM PSF classification
in each colour.

Likelihood is gauged by comparative density on the sky.  If, say, stellar-
PSF objects of R = B on annuli 5 arcsec from the set of all RASS X-ray
sources are 10 times as dense on the sky there compared with the all-sky
(background) density, then we say the chance of association of those
optical objects there is 90\%, i.e. of each 10 of those optical
objects, we take one as typical background and the excess 9 as causal.
This approach must incorporate local sky object density, as otherwise
calculated likelihoods in densely-populated areas would be falsely high
against the all-sky-average background.  A simple local density-dependent
multiplier would suffice in one sense, but this would overlook the
different mix of objects in different parts of the sky, i.e. the low-
density Galactic caps are expected to have a higher ratio of objects with
QSO-like colours than the high-density Galactic plane.  To accommodate
both density and object-mix variations, we have divided the sky into
twelve sky density bins, and accordingly have broken our optical catalogue
out into rectangles of approx 1 sq degree and allocated them by mean
object density into those twelve bins.  Table 5 shows the areas, object
counts, and average densities for the total objects and the APM-only
objects, for each sky density bin.  These density bins have been designed
to keep the discrepancy between any local sky density and the density of
the corresponding bin to a maximum of 20\%, although greater
discrepancies are possible in inhomogeneous areas, of course.  A 20 per
cent density error will result in a likelihood figure of e.g., 90 per
cent, to be written as 88\% or 92\%(see equation 2, below),
which we consider acceptable.

\begin{table*}
\caption{12 sky density bins and summations of the sky portion allocated
to each bin. Note that 243.71 square degrees are missing from the optical
catalogue.}
\begin{center}
\begin{tabular}{rrrrrrrr}
\hline
    Density &   Density range & Total area  &  Total no.  &   Mean &    APM Area
& APM no. &     APM mean\\
       bin  &   (per square degree) & (square degree) &     objects &
       density & (square degree) & objects &   density \\
\hline
       6000   &    1-- 6000 &  3206.24 &   15533000  &    4845  &   2757.65&
13057871   &    4735\\
       8000   & 6001--8000 &  5416.09  &  38352783   &   7081   &  4815.15&
33185681   &    6892\\
      10000   & 8001--10000 &  7333.54 &   65955475  &    8994  &   6382.29&
56028957   &    8779\\
      12000 &  10001--12000 &  6018.01 &   65589155  &   10899  &   4916.19&
52348586   &   10648\\
      15000 &  12001--15000 &  5591.52 &   74376431  &   13302  &   4039.37&
52316607   &   12952\\
      18000 &  15001--18000 &  3299.59 &   53680671  &   16269  &   1801.29&
28611437   &   15884\\
      22000 &  18001--22000 &  2409.98 &   46968715  &   19489  &    796.56&
15306870   &   19216\\
      34000 &  22001--34000 &  3539.80 &   94724199  &   26760  &    405.07&
10157923   &   25077\\
      45000 &  34001--45000 &  2380.68 &   93561653  &   39300  &     31.81&
1197891    &  37657\\
      60000 &  45001--60000 &  1144.29 &   56432307  &   49317  &     23.21&
1223691    &  52712\\
     100000 & 60001--100000 &   347.90 &   27239742  &   78298  &     32.18&
2529866    &  78608\\
     150000 &  over 100000  &  321.63  &  38511648   & 119740   &    16.54&
1970579   &  119123\\
\hline
      Total  &&              41009.25 &  670925779  &   16360  &  26017.32&
267935959  &    10298\\
\hline
\end{tabular}
\end{center}
\end{table*}

These binned areas and counts of objects serve as background
denominators for our likelihood calculations. For objects with APM PSF
information we use the APM areas and counts, for non-APM we use the
total areas and counts. One remaining division in our sky is that of
POSS-I versus UKST objects. As previously stated, UKST ($Bj - R$) is
0.65 of POSS-I ($O - E$) as a median, so an object typically will have a
larger colour spread in POSS-I than in UKST. Early pre-publication
versions of our catalogue calculated denominators separately for each
survey, thus doubling the number of bins and so reducing their
population. However, it is desirable to keep our background bin
populations as large as possible to minimize statistical fluctuations.
We judge that it is qualitatively preferable to use a simple
statistical rule to align the UKST colours to the POSS-I colours, thus
keeping these objects unified within the same bins. Thus we chose to
multiply each UKST object's $(Bj - R)$ by 1.5 ($\sim 1/0.65$) to map
to the statistically expected POSS-I $(O - E)$, for $Bj > R$. The
result is that the 12 sky density bins of Table 5 represent the
starting pools of data for all likelihood calculations. During each
such calculation, the appropriate pool was divided up by APM PSF class
and O-E colour to obtain the required background denominator.

Our APM-style PSF classification takes on just 4 discrete values for each
colour: stellar (written by us as `-' as a truncation of APM's `-1'),
fuzzy (`1'), extended (`2') and no classification (`n').  Our stellar and
fuzzy classes come straight from the APM, but our extended class `2'
differs from the APM merged-object `2' in that we expect that such a
source should have a visible source at the centroid, or be a component of
a large galaxy.  If the PSF is not classified as `-', `1', or `2', then we
take it as an `n' for these likelihood calculations even if the colour is
missing, as the question here is not the visibility but just the
morphology.  All objects are also accumulated into the PSF-free `n' class
in each colour (without double-counting if it is already `n'), and again
with `n' for both colours.  Thus, with just four PSF classifications
available for each of two colours, we have a total of 16 two-colour PSF
bins.

O-E colour is binned by 0.3 to keep bin populations large while blurring
colours by no more than 0.1 mag.  We use the range ($-0.9 \le O - E \le
4.5$), binned by 0.3, with $O-E < -0.9$ taken as -0.9 and $O-E > 4.5$ taken as
4.5.  As mentioned, for UKST $Bj > R$, we take $O-E = (Bj-R) \times 1.5$, then bin it
in the same way.  One-colour objects have no $O-E$, but are included in a
cumulation of all objects which is given a placeholder value of $O-E = 9.9$.
Thus we have a total of 20 $O-E$ colour bins.  Note that there is an APM
photometry artefact in dense LMC areas which results in an overabundance
of $Bj\ll R$ in the two highest density APM bins; possibly the APM confused
near neighbours when matching images across colours.  The consequence is
that we cannot use the colour criterion in the LMC.  Without this tool,
and in recognition that our methods are less effective in very dense
star fields, to deter false positives we have chosen to require 
co-positional fit within 1 arcsec to accept association in the two
highest density bins of 100000 and 150000.

The breakdown of our optical catalogue into these cross-categories of 12
sky density bins by 16 PSF bins by 20 colour bins is displayed at
http://quasars.org/docs/QORG-background.txt .  The total number and APM
number of objects for each of the 3840 cross-indexed bins are listed.  For
each likelihood calculation, a cross-indexed bin is selected using the
optical object's attributes, and that bin provides the background numbers
used for the denominator.

Likelihood is calculated in terms of the overabundance of optical objects
over the background.  As an example calculation, let us consider a HRI
source offset 3 arcsec from an optical object which is stellar in both
colours, has $O-E = 0.3$, and is located in sky of density bin 8000.  Our
input HRI catalogue has 6859 X-ray sources in sky of density bin 8000;
therefore for offset annuli of 3 arcsec about these, the total area
(between radii 2.5 and 3.5 arcsec) is 129,289 arcsec$^2$, and within this
area of sky our optical catalogue yields 31 objects (smoothed) which are
stellar in both colours and $O-E = 0.3$.  Table 5 shows that the all-sky
area of density bin 8000 is 4815.15 sq deg which converts to
62,404,324,852 arcsec$^2$, and within this sky area the background count of
objects which are stellar in both colours and $O-E = 0.3$ is 213,453, as
shown in `QORG-background.txt'.  The comparative sky density for these
optical objects at 3 arcsec offset from HRI sources is thus

\begin{eqnarray}
{\rm Density}   &=& ({\rm count }/{\rm area}) / ({\rm background\ 
     count}/{\rm background\ area})\cr
     &=&(31 / 129289{\ \rm arcsec}^2) / (213453 / 62404324852 {\ \rm 
     arcsec}^2)\cr
&=& 70.1
\end{eqnarray}

The density of 70.1 represents an overdensity of 69.1 compared to the
background of 1.  Thus confidence of association = $( 70.1 - 1 ) / 70.1 =
98.6$\% for each object, and this is our measure of causal
likelihood:

\begin{equation}
{\rm Confidence} = {({\rm density}-1)\over {\rm density}}
\end{equation}

Complete densities and supporting figures are given for all cross-indexed
bins for the HRI input catalogue in the density chart at
http://quasars.org/docs/QORG-HRI-densities.zip, and similarly for the RASS,
PSPC, WGA, NVSS, FIRST and SUMSS input catalogues.  Smoothing rules used
are itemized in the headers of those files.  Note that outlying bins such
as that of $O-E = -0.9$ can have very small populations, so to avoid small-
numbers fluctuations we have amalgamated the outliers to where the bin
population `count' in equation (1) is expected to be at least five.  Thus
in `QORG-HRI-densities.txt' the first displayed $O-E$ bin is $O-E = 0.3$,
which includes smaller $O-E$.  The need to keep bin populations high shows
that the efficacy of our likelihood method is directly dependent on the
size of the input catalogue, and indeed small-number fluctuations in
outlying bins are an occasional hazard.  In the closing section of this
paper we describe an offset-dependent penalty which we have deployed to
further control this intermittent problem.

There are, however, complications that we needed to resolve before these
final densities were written.  In the case of the X-ray catalogues, the
{\it ROSAT} fields are misaligned with respect to the optical background,
typically by 1-10 arcsec, and need to be shifted to their correct
locations.  Some shifting is also needed for the radio fields, but in this
case it is because the APM astrometry can be offset from the true by up to
2 arcsec in each of RA and DEC (at the plate edges; see MWHB for a full
discussion), and as we use the APM for our reference astrometry we need to
realign the radio survey astrometry where appropriate; that is, introduce
equal errors so as to align it to our APM background.  This is an
iterative process as a density chart must be compiled first out of the
original astrometry for each radio/X-ray catalogue, then that density
chart is used to re-align the astrometry, then a new density chart is
compiled using the revised astrometry as an input catalogue, etc.  Our
experience is that three iterations are sufficient as the fourth brings
little change to the density chart.  The final density charts are much
more focused than the initial ones, with high densities for near
positional fits, and densities falling off rapidly outwards, much like the
final chart displayed on http://quasars.org/docs/QORG-HRI-densities.zip for
HRI; similar results are obtained for the other catalogues. We describe
our method for achieving these shifts in the following sections.

\subsection{The X-ray Sources}

The immediate consideration in using {\it ROSAT} X-ray catalogue data
is in deciding which source detections to use at all, as their
reliability varies and flags are present to signal reduction
difficulties due to close or complex sources. Most HRI and PSPC
sources bear some of these flags; of the 131,902 total HRI sources,
only 13,452 are entirely unflagged. These flags originate from the
surveyors' manual inspection of all the individual detections, and one
of the flags signals their overall assessment that the source is a
false detection; where this flag is not set, the source was not
determined to be spurious. We therefore use onwards all sources
without this flag as candidates for matching to our optical catalogue.
Of the 131,902 HRI sources, 111,865 are without the false-detection
flag; however, of these, 8767 are astrometric duplicates (to the
1-arcsec resolution of this project) within the same {\it ROSAT}
observing field, and 46,700 further sources are flagged by HRI as
`non-unique' astrometric duplicates across different {\it ROSAT}
fields -- this is not unexpected, as many objects of interest were
observed repeatedly. Thus in the end we are left with 56,398
astrometrically unique HRI sources to attempt to match to optical
objects. Similarly, 100,205 individual PSPC sources are available to
us from the 118,785 original sources in the combined PSPC and PSPCF
catalogues; these catalogues have no `non-unique' flag. The WGA catalogue
authors used a single `quality flag' to gauge reliability, and using
their 88,621-record catalogue of `good' sources yields 88,378
individual sources. The RASS catalogue has clean data with only a few
complex-emission sources which we have chosen to retain, so we use
their full complement of 124,730 sources.

The primary task in associating {\it ROSAT} sources with optical
objects is that of astrometrically fitting the {\it ROSAT} observing
fields to the optical background. As detailed in Appendices B and D of
the {\it ROSAT} User's Handbook, there were ongoing boresighting and
undiagnosed errors which caused pointing unreliability of up to 20
arcsec. This `attitude solution error' was accompanied by a systematic
roll angle error of 6 arcsec which has been corrected for in the final
HRI, PSPC and RASS catalogues that we use, but the attitude error was
more random than systematic and persisted throughout {\it ROSAT}'s
operation. HRI fields are nominally more precisely pointed than PSPC
or RASS, but we find in our analysis (below) that some HRI fields,
too, are offset by as much as 15 arcsec; see also Mason et al (2000),
figure 1, which shows PSPC sources offset from their optical
counterparts by up to 15 arcsec with one source offset by 30 arcsec.
WGA fields often have offsets 10 arcsec greater than their
corresponding PSPC fields, possibly because of the absence of the roll
angle fix combined with an early pointing solution. The question of
correctly repointing a {\it ROSAT} observing field is present in every
instance of its use. Researchers have often resisted shifting the
fields lest their analysis be disputed. Our task here, however,
explicitly involves causal linking of optical and X-ray sources, and
correctly repointing the {\it ROSAT} fields is essential to optimizing
this task. We believe our likelihood algorithms based on our whole-sky
optical data gives us an unprecedented opportunity to decide the
correct alignment of the {\it ROSAT} fields in bulk.

The general principle of our approach is to find compelling X-ray-optical
associations and shift each {\it ROSAT} field so as to superpose its X-ray
sources perfectly onto the optical background.  Of course, the real data
never fits perfectly, many X-ray sources have optical counterparts too
faint for our optical catalogue, and we need to find quantifications which
yield optimal alignments without falling prey to chance coincidental fits.
Our main tool is of course the likelihood confidence method explained in
the previous section, and we needed to determine the likelihood score for
each X-ray-optical association and sum the scores for each {\it ROSAT} field in
a way which incorporates both (1) the number of associations and (2) the
power of precise fitting associations in a balanced way -- neither of
these is sufficient on its own, as random alignments can easily give rise
to many associations at large offsets, or a few small-offset associations.
Monte Carlo simulations cannot easily be designed to optimise the
combination of these two measures, as we have no a priori notions of what
comparative configurations of control and test data should be expected to
fit validly, and which would fit only coincidentally.  Any simulation-
derived rules would need to be tested against real-sky data to find if the
simulation was designed in conformance to real-sky behaviour; the
requirement for real-sky testing renders the simulation superfluous.  Our
general approach of being guided by the real data itself to find the rules
and numbers applied as strongly here as anywhere else.  Thus, in practice,
to determine the optimal combination of the above two measures, we
heuristically tried different formulations and tested them against well-
understood X-ray fields to find the best-performing solution.

We processed each input catalogue (e.g., HRI) separately. Our first
step was to compile an initial density chart (as defined in the
preceding section) for a whole input catalogue using its nominal
(original) {\it ROSAT} astrometry. Next we test, for each {\it ROSAT}
field, all positional shifts from the nominal location out to $\pm 48$
arcsec offset in each of RA and DEC (in intervals of 3 arcsec to save
processing, thus 1089 shifts in total). Each tested field shift is
scored as follows: first, we use the density chart to produce
confidence of association figures for the field's X-ray- optical
matches, using sources singly only; these need to be amalgamated into
a final score for that field shift. This final score must incorporate
both the number of X-ray-optical matches and their individual
confidence scores; thus a summation of confidence scores is indicated,
but in testing this against selected fields (notably the quasar-rich
environs of NGC 3628) we found that field shifts with many
low-confidence matches tended to outscore field shifts with a few
high-confidence matches which were in fact correct, judged by
co-positionality of X-ray sources to known quasars. We found this
problem to be remedied by using the squares of the confidence scores
instead of the confidence scores directly; in this way a single 100\%
match is worth four 50\% matches instead of just two. This yielded the
correct final astrometry in our test fields. It did not, however, work
to use the cubes (etc) of the confidence scores, as then a single
randomly-generated precise co-positionality could overpower a small
number of valid causal matches. Thus in our summations we define the
`weight' of an individual X-ray-optical match to be the square of its
confidence figure. We double a weight figure if its optical object is
a known QSO, and decrease it up to 33\% where the optical astrometry
is compromised due to non-stellar morphology or missing $R$ or $B$;
again, these corrections evolved heuristically via extensive testing.
Only individual weight figures of $>0.5$, corresponding to confidences
of $>70$\%, are retained to limit the contributions of random
matches, and at least two separate X-ray-optical associations must be
present for a field shift to be plausibly informative; to shift a
field based on a single association prejudges the process. The total
weight (score) of the field shift is the sum of the weights of all its
individual X-ray-optical combinations.

Of course, the significance of this score depends on the number of X-ray
sources in the {\it ROSAT} field, which we term $N$.  Finding two precise X-ray-
optical alignments in a {\it ROSAT} field having only two X-ray sources might
constitute a compelling field shift, but if the field has 100 X-ray
sources, and we have matched only two, then that would be unconvincing.
We need to make this quantitative.  One might start by considering the
contribution of the field's angular size and photometric depth to $N$, but
various studies (including Mason \etal\ 2000) have found that the
associability of X-ray signatures with visible optical sources does not
vary much with X-ray flux.  Thus we can quantify $N$ directly as the sole
counterbalance to our total weight of the field shift; it is the sole
counterbalance because on the optical side our density calculations
already incorporate the optical object density.  We incorporate this
quantification $f(N)$ to define the `power' $P$ of the field shift:

\[
P = \Sigma ({\rm weight}) \times f(N)
\]

We will find a threshold power value below which we deem that the field
shift is not proven and so not used.  We find a suitable $f(N)$ by heuristic
testing.  $f(N)=1/N$ fails because it models match numbers to be increasing
linearly with X-ray source numbers, ignoring high match rates randomly
obtained in low-density fields, i.e. small-numbers fluctuations; we find
that twice the matches in a field with twice the X-ray detections is
indeed more significant as our testing shows such field-shifts point more
reliably to known quasars.  $f(N)=1/{\sqrt N}$ is found to model adequately
the performance of field shifts using fields of different $N$; this is again
a heuristically-gained measure.  Thus we define the power of the X-ray-
optical alignment of the field as:

\begin{equation}
P = \Sigma ({\rm weight}) / \sqrt{N}
\end{equation}

where $N$ is the number of X-ray sources in the {\it ROSAT} field.
Note that we thus ascribe the same power to two precise superpositions
in a field of four X-ray sources as to 10 precise superpositions in a
field of 100 X-ray sources, where other X-ray sources are unmatched to
optical. This conforms to our na\"\i ve expectation. Large fields with
many X-ray sources, such as some RASS fields, will have high power
scores only if they are well aligned with the optical background.

However, yet another factor is needed to counteract high weight scores
generated randomly at large field shifts; after all, two configurations of
random point sources will align optimally, but meaninglessly, somewhere,
most likely at large field shifts as the number of candidate field shifts
increases linearly with shift distance.  Thus, we need an accompanying
linearly-dependent penalty to suppress the random outliers.  The question
basically is that of quantifying the significance of the original
astrometry as specified in the {\it ROSAT} catalogues.  We analyse this by
compiling the mean power score of all {\it ROSAT} fields over all 1089 candidate
shifts, for each {\it ROSAT} catalogue, in Table 6.  Inspection shows the mean
power rating is highest at the original astrometry and falls off with
increasing field shift until at high shifts it stabilizes into a
background level.  The HRI mean power at the original astrometry is less
than that of the PSPC because its smaller field sizes provide fewer
associations per field.  The rapid dropoff of the HRI mean power with
increasing field shift shows that it is the best-pointed of the four
surveys, and its reaching near-stability at a shift of 21 arcsec indicates
that there are no valid HRI field shifts greater than 20 arcsec.  The PSPC
and WGA powers decline significantly out to about 30 arcsec.  The WGA mean
power at the original astrometry is small because of its lesser pointing
accuracy; in addition, about 30\% of the listed WGA fields are
amalgamations of multiple PSPC fields for which our field shifting
technique is necessarily problematic.  The RASS powers in Table 6 are
medians as we process RASS differently to the others; we describe this in
more detail below.

\begin{table*}
\caption{Mean power and total number of $>70$\% confidence
associations for all {\it ROSAT} fields for each catalogue, for
candidate shift increments of 3 arcsec. Object numbers increase at
higher shifts because of the greater quantity of candidate field
shifts. RASS powers are medians.}
\begin{center}
\begin{tabular}{rrrrrrrrr}
\hline
Field shift &   \multicolumn{2}{c}{HRI} & \multicolumn{2}{c}{PSPC}  &
\multicolumn{2}{c}{WGA} & \multicolumn{2}{c}{RASS}\\
(arcsec)  &    Power  &  No. objects  &   Power  &  No. objects  &
Power  &  No. objects & Power  &   No. objects\\
\hline
    0  &     1.38  &     2251  &    1.61  &     3220  &     1.07  &     1792&
0.93  &     1144\\
    3  &     1.22  &    16048  &    1.52  &    25010  &     1.05  &    13960&
0.89  &     9082\\
    6  &     0.96  &    19406  &    1.37  &    35830  &     1.02  &    19944&
0.79  &    13537\\
    9  &     0.73  &    19750  &    1.20  &    44874  &     0.98  &    24940&
0.67  &    17864\\
   12  &     0.58  &    29180  &    1.01  &    81977  &     0.92  &    45276&
0.54  &    35319\\
   15  &     0.51  &    20527  &    0.83  &    64890  &     0.84  &    35261&
0.43  &    30367\\
   18  &     0.49  &    26240  &    0.72  &    83211  &     0.76  &    44561&
0.36  &    42535\\
   21  &     0.48  &    24849  &    0.63  &    75432  &     0.69  &    39375&
0.31  &    41769\\
   24  &     0.48  &    29044  &    0.59  &    83235  &     0.62  &    41781&
0.27  &    49226\\
   27  &     0.47  &    40676  &    0.56  &   109828  &     0.57  &    52335&
0.25  &    68633\\
   30  &     0.47  &    33109  &    0.54  &    86150  &     0.53  &    38722&
0.23  &    55324\\
   33  &     0.47  &    42010  &    0.53  &   106248  &     0.52  &    45635&
0.22  &    69625\\
   36  &     0.47  &    39520  &    0.52  &    97874  &     0.50  &    40444&
0.21  &    64784\\
   39  &     0.47  &    50532  &    0.51  &   123597  &     0.50  &    49278&
0.20  &    82234\\
   42  &     0.47  &    50237  &    0.51  &   120733  &     0.50  &    46913&
0.19  &    80655\\
   45  &     0.47  &    46927  &    0.50  &   113371  &     0.49  &    42863&
0.19  &    75379\\
   48  &     0.47  &    61338  &    0.49  &   146744  &     0.49  &    53524&
0.18  &    97355\\
\hline
\end{tabular}
\end{center}
\end{table*}

We used the power values in Table 6 as our measure of the significance
of the nominal astrometry of the four {\it ROSAT} catalogues, to be
added to the power score of each field shift, thus favouring lesser
shifts where all else is equal. Before we added this in, though, we
analysed the full set of 1089 candidate shifts for each field to find
local power maxima; i.e., field shifts having power values higher than
all their neighbouring shifts have, which generally signifies close
individual alignments across the X-ray and optical fields. We might
find, say, 38 of these, and we use from then on only those 38, which
thus avoids skewing positions when adding the values from Table 6. We
then added the extra power score from Table 6 according to the field
shift in arcsec for each candidate shift, but also subtract the score
obtained for no offset zero (e.g. 1.38 for HRI) to normalize the score
compared to non-shifting fields; the final effect is that of a penalty
against the original astrometry, i.e. the further the candidate shift,
the greater the penalty deducted from that candidate's power score.
After applying this penalty, the field shift with the highest total
power score exceeding the threshold value of 0.5 is the `winning'
field shift, and is used from then on, provided it leads the runner-up
power score by at least 0.1 or if both shifts are astrometrically
similar -- we prefer to use no shift if the top shift candidates are
scoring about the same, as can happen especially in dense star fields
where random fits often have equally `good' power scores. The 0.5
($\ge0.45$) power threshold was found by trial and error and
physically corresponds to two 70\% confidence associations in a
field of four X-ray sources; `best' fields scoring less than this
usually look like random fits. The 0.1 power distinction approximately
corresponds to the presence of an additional 70\% confidence
association. As the winning field shift was selected from candidates
at intervals of 3 arcsec, we tested further field shifts offset 1
arcsec from the winner to find the one producing the best score; this
is the final field shift used. A complete list of the HRI fields and
the field shifts used is displayed in
http://quasars.org/docs/HRI-fields.txt, and similarly for the RASS,
PSPC and WGA input catalogues.

We have established maximum shift values of 18 arcsec for HRI and PSPC
and 31 arcsec for WGA. These were not arbitrary decisions but were
made after an initial full build without using these maxima, and
without using the astrometric significance penalties from Table 6. See
http://quasars.org/docs/HRI-shifts-old.ps for the distribution of HRI
field shift distances versus power; each field is represented once, by
the shift of that field that yields the best power score; similar
charts are available for RASS, PSPC and WGA. The graph shows a
population along the vertical power axis consisting of high-power
($>1.5$) low-distance ($<10$ arcsec) shifts, and another population
along the horizontal shift distance axis consisting of high-distance
($>15$ arcsec) shifts of low power ($<1.5$); these are the
randomly-generated field shifts which have no physical significance
and arise only because of the sheer volume of high-distance candidate
field shifts. We originally tried to draw a dividing line of
significance where these two populations meet, which is of course not
a clean boundary as valid and invalid shifts are found on either side.
Spot checks of all fields in the central vicinity where the dividing
line lay revealed that beyond a certain maximum field shift no shift
looked compellingly good; either fuzzy or one-colour objects dominated
or there were a lack of close positional fits. For HRI we found the
maximum good field shift was 18 arcsec and for PSPC we found the same;
although we felt PSPC should have some larger good field shifts, given
the intrinsically lower resolution of the observations, we could find
no compelling instance in our extensive spot checks. The valid-looking
shifts all had good power scores, and shifts of similar magnitudes
with low power scores looked less compelling on inspection. These
low-power shifts are generally removed by the astrometric significance
penalty from Table 6. WGA had plausible alignments out to a 31 arcsec
shift, and a broad view of http://quasars.org/docs/WGA-shifts.ps shows
that as a whole the WGA fields are more free-ranging than HRI or PSPC.
Having established these maximal field shift values for HRI, PSPC and
WGA, the final full build was done which disallowed consideration of
any fields shifts beyond the maxima, and which required any candidate
field shift to have a power score $\ge 0.45$ above the astrometric
significance penalty from Table 6, as described above. The result for
HRI is shown on http://quasars.org/docs/HRI-shifts.ps and similarly for
the PSPC and WGA input catalogues.

The RASS differs from the other {\it ROSAT} surveys in that its fields
are large ($\sim 27$ deg$^2$ each) and the exposures comparatively
short, with concomitant large uncertainties in the published source
positions. We have also encountered astrometric inconsistencies within
RASS fields which are possibly due to distortion in the outer off-axis
parts of the {\it ROSAT} images. Given this graininess of the RASS
positions, we have elected to optimize our optical selections by using
the HRI and PSPC surveys to `anchor' the RASS fields wherever
possible, by correlating high-flux X-ray sources across the three
catalogues and designating the corresponding HRI/PSPC-chosen optical
objects as highly-weighted targets for the RASS fields. RASS fields
without HRI/PSPC overlaps must still rely on astrometric significance
penalties to avoid randomly-large shifts, and we find that
median-based power values accord best with the grainy RASS astrometry
to allow valid-looking large shifts to be selected. A `valid- looking
large shift' is one for which associated optical objects have similar
PSFs and colours as those associated in fields with small shifts,
and which contains some close X-ray-optical positional fits. We found
that some large RASS field offsets did fulfil these criteria, so we
did not impose a maximum shift value as was done with HRI and PSPC.
However, even without such a limit there turn out to be few large RASS
field shifts, as seen on http://quasars.org/docs/RASS-shifts.ps . We
have checked all fields with shifts of $>14$ arcsec: fields 33023034 at
42 arcsec and 33016040 at 34 arcsec are the two largest shifted fields,
and both have multiple good optical fits and sources confirmed by
PSPC. All the other fields also look valid except for three low-power
fields which looked like random `best' fits: 33025019 at 15 arcsec and
0.7 power, 33012017 at 15 arcsec and 0.9 power and 33031016 at 26
arcsec and 1.1 power. We have manually reset these to zero shift and
none contributes any associations to the final catalogue. Having
culled these, we are satisfied with the performance of the large RASS
field shifts.

Table 7 summarizes all field shifts for the four input {\it ROSAT} catalogues,
showing the resultant increase in the number of $>70$\% confidence X-
ray-optical associations.  For HRI the number of associations presented in
QORG is less than the number of $>70$\% confidence associations used
to shift the fields; this is because of overlapping-field duplicates which
we remove; WGA has few such duplicates, and RASS none.  The shift=0 row
represent fields which were `shifted' to their original locations; the
lack of astrometric penalty at zero shift allows a few low-quality fields
to reside there.  Unshiftable fields are included for completeness as
`unshifted'.  The high number of WGA fields without a preferred shift is a
consequence of their $\sim 1000$ merged fields which cause problems for our
analysis, and many HRI fields are left unshifted because they contain few
sources; 3288 HRI fields have fewer than 10 sources, compared to 2005 for
PSPC.

In all, for shifted fields, HRI shows the high-confidence ($\sim
88$\%) X-ray-optical associations expected from their well-pointed
high-resolution observations, PSPC's pointing looks as good but its
detections are not as well resolved so positional fluctuations lower
the median confidence of X-ray-optical associations to about 79 per
cent, WGA's resolution is the same as PSPC's but has pointing problems
which lower the median confidence of X-ray-optical association to
about 70\%, and RASS's pointing is similar to PSPC's but its
resolution appears to be quite grainy with X-ray-optical offsets often
in excess of the stated positional uncertainty, which keeps the median
confidence of X-ray-optical associations down to about 69\%.
Table 8 displays the median offset between the original published
X-ray position and any optical object which we find to be associated
with $>40$\% confidence (which is the threshold required for inclusion
in QORG), categorized by published positional uncertainty of the X-ray
detection; {\it ROSAT} duplicate entries are included, and we use the
original astrometry to exclude the effect of our field shifting. No
RASS sources are published with less than 6 arcsec positional
uncertainty. It can be seen that HRI has marginally better accuracy
than PSPC, which is in turn marginally better pointed than RASS, which
shows the RASS positions to have greater scatter and thus a lower
resolution. There is no WGA entry in Table 8 as WGA provides no
published positional uncertainty for their detections; their published
comment is that the uncertainty is `close to 10 arcsec' which accords
well with our finding that the median WGA X-ray-optical offset is 8
arcsec regardless of source flux.

\begin{table*}
\caption{Alignment of {\it ROSAT} fields: The number of fields shifted
as a function of shift distance, with X-ray source numbers, $>70$ per
cent-confidence associations used in shifting (both before and after
the shift), resultant associations appearing in our catalogue, and the
median confidence of those associations. Unshifted fields are
included; the absence of a shift is generally
due to a lack of good X-ray-optical fits.}
\begin{tabular}{rrrrrrrrrrrrrr}
\hline
&\multicolumn{6}{c}{HRI}&\multicolumn{6}{c}{PSPC}\\
 Shift  &  No.  &  Orig. & $>70$\% & QORG &No. in &Median & No. & Orig
 &$>70$\% & QORG& No. in &Median\\
 (arcsec) & fields &sources& conf& $>70$\%conf& QORG & conf. & fields& sources&
 conf& $>70$\% conf& QORG  &conf\\
\hline
   0  &     58  &   961  &    324  &    324  &   270  &  80  &    90  &  2577&
948  &     948  &    839  &  77\\
   1  &    301  &  3945  &   1608  &   1692  &  1276  &  88  &   285  &  7576&
2679  &    2779  &   2824  &  78\\
    2  &    354  &  4334  &   1800  &   2122  &  1646  &  89  &   324  &  8920&
3008  &    3412  &   3231  &  78\\
    3  &    480  &  7331  &   2345  &   3022  &  2549  &  88  &   575 &  14355&
4628  &    5481  &   5093  &  78\\
    4  &    456  &  6090  &   1889  &   2708  &  2240  &  88  &   558 &  13650&
4186  &    5206  &   4980  &  78\\
    5  &    271  &  3515  &    993  &   1557  &  1301  &  88  &   316  &  7915&
2491  &    3160  &   2903  &  79\\
    6  &    131  &  1427  &    381  &    712  &   530  &  87  &   274  &  6177&
1767  &    2386  &   2311  &  78\\
    7  &     92  &  1075  &    247  &    468  &   417  &  84  &   221  &  4981&
1400  &    2020  &   1887  &  78\\
    8  &     27  &   209  &     53  &    113  &    85  &  89  &   108  &  2232&
581  &     884  &    802  &  78\\
    9  &     20  &   182  &     45  &     87  &    72  &  89  &   103  &  1929&
517  &     851  &    664  &  79\\
   10    &    3  &    33    &    6  &     15  &    15  &  81  &    46  &   808&
203  &     357  &    276  &  77\\
   11    &    4  &    20    &    3  &     16  &     8  &  81  &    27  &   460&
115  &     207  &    177  &  82\\
   12    &    1  &     6    &    1    &    4  &     5  &  84  &    13  &   172&
39    &    84  &     75  &  82\\
   13    &    .  &     .    &    .    &    .  &     .  &   .  &    13  &   248&
42    &    99  &     59  &  74\\
   14    &    1  &    11    &    1    &    8  &     9  &  92  &     7  &   288&
46    &    99  &     82  &  82\\
   15    &    .  &     .    &    .    &    .  &     .  &   .  &     7  &   154&
32    &    60  &     66  &  79\\
   16    &    .  &     .    &    .    &    .  &     .  &   .  &     3  &    37&
6    &    20  &     18  &  81\\
   17    &    1  &    33    &    6  &     13  &    12  &  79  &     1  &    16&
3    &     8    &    8  &  89\\
   18    &    1  &    10    &    .    &    6  &     4  &  88  &     1  &     8&
1    &     3    &    3  &  87\\
unshifted & 2920 &  27121  &   1251  &   1251  &  2294  &  60  &  2321 &  29073&
1651  &    1651  &   3174  &  58\\
\hline
Total  &   5121  & 56303  &  10953  &  14118 &  12733  &  84  &  5293 & 101576&
24343  &   29715  &  29472  &  76\\
\hline
\\[10pt]
\hline
&\multicolumn{6}{c}{WGA}&\multicolumn{6}{c}{RASS}\\
 Shift  &  No.  &  Orig. & $>70$\% & QORG &No. in &Median & No. & Orig
 &$>70$\% & QORG& No. in &Median\\
 (arcsec) & fields &sources& conf& $>70$\%conf& QORG & conf. & fields& sources&
 conf& $>70$\% conf& QORG  &conf\\
\hline
    0  &    14  &  520  &    92  &    92  &  149   & 66  &   41 &   3672&
665  &    665  &  1022  &  71\\
    1  &    41  &  1187  &   264  &   270  &  342  &  70  &   71 &   6372&
1213  &   1247  &  1876  &  70\\
    2  &    63  &  1862  &   352  &   405  &  574  &  68  &   88 &   9683&
1711  &   1863  &  2935  &  69\\
    3  &   114  &  3872  &   707  &   818  &  1099 &   67  &  183 &  18954&
2886  &   3213  &  4976  &  68\\
    4  &   150  &  4531  &   823  &   991  &  1284 &   68  &  187 &  19905&
2961  &   3383  &  5417  &  67\\
    5  &   100  &  2733  &   501  &   655  &  823  &  70  &   87 &   8081&
1440  &   1694  &  2522  &  71\\
    6  &   124  &  3276  &   622  &   813 &   1033  &  71  &  107 &   9156&
1606  &   1925  &  2862  &  69\\
    7  &   152  &  4553  &   814  &  1103  &  1438  &  70  &  102 &   9134&
1415  &   1687  &  2629  &  69\\
    8  &    97  &  2714  &   482  &   682  &  796  &  72  &   41 &   3523&
568  &    725  &  1044  &  69\\
    9  &   131  &  3636  &   553  &   874 &   1109  &  71  &   54 &   4539&
759  &    962  &  1461  &  70\\
   10  &    94  &  2779  &   418  &   639  &  824  &  70  &   29 &   2459&
334  &    441  &   665  &  67\\
   11  &    88  &  2493  &   377  &   620  &  761  &  69  &   18 &   1461&
209  &    290  &   413  &  69\\
   12  &    78  &  2485  &   319  &   617  &  731  &  71  &   11  &  764&
86  &    121  &   183  &  68\\
   13  &    40  &  1112  &   134  &   279  &  335  &  71  &   10  &  727&
117  &    158  &   226  &  67\\
   14  &    50  &  1304  &   167  &   351  &  462  &  71  &    2  &  149&
14  &     28  &    43  &  71\\
   15  &    45  &  1420  &   182  &   346  &  463  &  70  &    4  &  413&
49  &     51  &    70  &  68\\
   16  &    32  &  729  &    92  &   207  &  225   & 74  &    2  &  152&
18  &     21  &    36  &  75\\
   17  &    16  &  452  &    57  &   124  &  128   & 73  &    .  &    .&
  &      .  &     .  &  .\\
   18  &    11  &  225  &    21  &    62  &   70  &  79  &    .  &    .&
  &      .  &     .  &  .\\
   19  &    13  &  317  &    40  &    99  &   98  &  77  &    3  &  167&
4  &     14  &    36 &   58\\
   20  &     9  &  233  &    30  &    73  &   61  &  75  &    1  &   90&
3  &      6  &    15  &  49\\
   21  &     7  &  130  &    19  &    54  &   29  &  69  &    2  &   80&
4  &      8  &    11  &  63\\
   22  &     8  &  236  &    17  &    62  &   87  &  73  &    .  &    .&
  &      .  &     .  &  .\\
   23  &     6  &   74  &    11  &    27  &   20  &  77  &    .  &    .&
  &      .  &     .  &  .\\
   24  &     5  &  134  &    11  &    38  &   49  &  70  &    .  &    .&
  &      .  &     .  &  .\\
   25+  &   18  &  324  &    33  &   103  &   66  &  72  &    2  &  128&
12  &     16  &    23 &   73\\
unshifted & 2479 &  44942  &  2086  &  2086  &  5656  &  56  &  332  & 25067&
737  &    737  &  2056  &  55\\
\hline
 Total  &  3985 &  88273  &  9224  &  12490 &  18712  &  65  &  1377 & 124676&
16811  &  19255  &  30521  &  68\\
\hline
\end{tabular}
\end{table*}

\begin{table*}
\caption{The median offsets (in arcsec) from the original {\it ROSAT}
coordinates to a $>40$\%-associated optical object, by published
positional uncertainty (in arcsec) of the X-ray source.  The median
offsets correspond linearly to scatter and inversely to resolution,
showing HRI to have the best astrometric accuracy, followed by PSPC.}
\begin{center}
\begin{tabular}{rrrrrrr}
\hline
Positional  &  HRI med  &  HRI no. &   PSPC med &  PSPC No.  &   RASS med  &  RASS No.\\
uncertainty &   offset &  sources &   offset &   sources  &   offset    &   sources\\
\hline
   0-1    &       3    &    2758    &    4    &    2095    &     .    &        .\\
    2    &        4    &    5104    &    5    &    2319    &     .    &        .\\
    3    &        4    &    1635    &    5    &    3097    &     .    &        .\\
    4    &        5    &    1069    &    5    &    3718    &     .    &        .\\
    5    &        5    &     821    &    5    &    2696    &     .    &        .\\
    6    &        5    &     386    &    6    &    3682    &     5    &      126\\
    7    &        5    &     255    &    6    &    2981    &     5    &      581\\
    8    &        6    &     236    &    6    &    2453    &     6    &     1312\\
    9    &        6    &     200    &    6    &    1582    &     6    &     1384\\
   10    &        6    &     185    &    7    &    1576    &     7    &     1826\\
 over 10    &     6    &     335    &    7    &    3365    &     8    &    24613\\
\hline
 total    &       4    &   12984    &    5    &   29564    &     7    &    29842\\
\hline
\end{tabular}
\end{center}
\end{table*}

In the end, the question of justification remains; that is, do our
field shifts indeed correctly align the {\it ROSAT} fields with the
optical background? As a final check we were able to use the
recently-published catalogues from the First {\it XMM-Newton}
Serendipitous Source Catalogue (XMM1, 2003) and the ChaMP First X-ray
Source Catalog (Kim \etal\ 2003) to verify our optical selections.
These catalogues derive their detections from the high-resolution {\it
XMM-Newton} and {\it Chandra} satellite observatories which are the
next generation after {\it ROSAT}. Their nominal positional errors are
typically 1-4 arcsec depending on source flux, and where possible they
each use astrometric solutions against the optical background to hone
their astrometry by a few arcsec; in this they share our premise that
such optical matching is an appropriate tool. The XMM1 catalogue
contains 41,990 good-to-medium quality detections, representing about
36,000 unique sources, which we mapped to 12,423 unique objects in our
optical catalogue using a matching radius of 5 arcsec. The ChaMP
catalogue is much smaller with just 991 detections representing 974
unique sources which we have mapped to 379 objects in our optical
catalogue using the same method. It was necessary, before the main
test, to match the XMM1 and ChaMP catalogues against each other to see
how well they agree. We found 86 X-ray sources in common between the
two catalogues, of which 80 were placed within 5 arcsec of each other;
this accords well with the nominal positional error of 4 arcsec, and
the outliers (out to an 11 arcsec discrepancy) hail from star-poor
areas where optical astrometric solutions were not used. We searched
for optical astrometric matches to these shared X-ray sources within 2
arcsec of the listed X-ray positional error, the 2 additional arcsec
accommodating both rounding and the 1 arcsec error typical of our
optical catalogue; we call these `good' matches. Using this matching
criterion we found that 29 of these shared sources map on both sides to
objects in our optical catalogue. All but one of these shared
detections agreed on the optical object selected, which yields an
optical hit ratio of 98\% (57/58), assuming the joint optical
associations are all true sources. Although we are here in the realm
of small numbers, the consistency between the two catalogues encouraged
us to consider XMM1 and ChaMP optical associations to be reliable
tests of the accuracy of our optical selections for the {\it ROSAT}
detections. The comparison of the XMM1 and ChaMP joint detections is
viewable at http://quasars.org/docs/XMM1-vs-ChaMP.txt .

We matched the {\it ROSAT} sources unambiguously to the XMM1 and ChaMP sources
by finding unique X-ray source matches within 30 arcsec radii which have
similar normalized fluxes, i.e. the stronger flux is less than twice the
other.  The `good' optical matches to these XMM1/ChaMP sources gave us
precise optical targets against which to measure the performance of our
field-shifted positions compared with the original {\it ROSAT} astrometry.  This
is a very precise test, as the optical targets and {\it ROSAT} positions, both
original and shifted, are all specified to arcsec precision on our optical
background; {\it ROSAT} positional uncertainties are immaterial as we are
testing catalogued positions, not true source positions.  This test is
viewable on a case-by-case basis at http://quasars.org/docs/QORG-vs-Original-ROSAT.txt and is summarized in Table 9 which displays simple
counts of X-ray-optical associations as a function of offset in arcsec for
each of the four {\it ROSAT} catalogues.  The accumulator columns of Table 9
(labelled `Total') show that our catalogue has twice (158/75) the accuracy
of the original HRI catalogue in pinpointing correct optical sources
within offsets of 2 arcsec inclusive and maintains a robust advantage out
to 5 arcsec (280/228), after which the numbers even out, as expected.
Gains are modest with PSPC, with just a 36\% (166/122) advantage
within offsets of 2 arcsec and just 13\% (562/498) to 6 arcsec.
Gains are very good with WGA, with twice (162/87) the capture rate within
offsets of 4 arcsec and still strong (278/194) to 6 arcsec.  And with RASS
we start well with a 75\% (42/24) advantage within offsets of 5
arcsec but it evens out rapidly beyond that.  Overall we are pleased with
the performance of our field shifts against the HRI and WGA catalogues,
whilst a little disappointed that our improvements against PSPC and RASS
are not equally strong; perhaps off-axis vignetting and blurring
(documented on pages 20-23 of the {\it ROSAT} User's Handbook) in the outer
parts of large-field {\it ROSAT} exposures resulted in astrometric distortion
which would cause problems for our method.

\begin{table*}
\caption{Performance of QORG shifted source locations compared with
original {\it ROSAT} source locations when tested against optical
targets identified by XMM1/ChaMP sources. For each offset in arcsec,
the number of X-ray/optical pairings found is listed for shifted QORG
fields and original {\it ROSAT} fields in turn. The `Total' columns
are running totals of the `No.' columns. All four {\it ROSAT}
catalogues are represented.}

\begin{tabular}{lrrrrrrrrrrrrrrrr}
\hline
Opt/Xray&\multicolumn{4}{c}{HRI}&\multicolumn{4}{c}{PSPC}&\multicolumn{4}{c}{WGA}&\multicolumn{4}{c}{RASS}\\
offset & QORG & Orig & QORG & Orig & QORG & Orig & QORG&  Orig  &QORG & Orig  &QORG&Orig&  QORG & Orig&  QORG & Orig\\
(arcsec)  & No.  &  No. &  Total &  Total  &  No.  &  No.  & Total &  Total  &  No.  &  No.  & Total&
Total  &  No.  &  No.   Total  & Total\\
\hline
0  &  19  &  5  & 19  &  5  & 16  &  4  & 16  &  4  & 12  &  0  & 12&
0  &  7  &  0  &  7  &  0\\
1  &  67  & 30  & 86  & 35  & 51  & 41  & 67  & 45  & 18  & 17  & 30&
17  &  3  &  2  & 10  &  2\\
  2  &  72  & 40 & 158  & 75  & 99  & 77 & 166 & 122  & 37  & 16  & 67&
33  &  7  &  5  & 17  &  7\\
  3  &  58  & 65 & 216 & 140  & 97  & 94 & 263 & 216  & 38  & 23 & 105&
56  &  5  &  9  & 22  & 16\\
  4  &  40  & 53 & 256 & 193 & 105 & 109 & 368 & 325  & 57  & 31 & 162&
87  &  6  &  4  & 28  & 20\\
  5  &  24  & 35 & 280 & 228 & 103  & 92 & 471 & 417  & 63  & 58 & 225&
145  & 14  &  4  & 42  & 24\\
  6   & 9  & 44 & 289 & 272  & 91  & 81 & 562 & 498  & 53  & 49 & 278&
194  &  7  & 14  & 49  & 38\\
  7  &  11  & 18 & 300 & 290  & 53  & 71 & 615 & 569  & 46  & 55 & 324&
249  &  6  & 10  & 55  & 48\\
  8   & 7  &  9 & 307 & 299  & 48  & 58 & 663 & 627  & 40  & 48 & 364&
297  & 10  & 11  & 65  & 59\\
  9   & 5  & 13 & 312 & 312  & 41  & 50 & 704 & 677  & 35  & 43 & 399&
340  &  6  &  9  & 71  & 68\\
  10   & 2  &  3 & 314 & 315  & 40  & 41 & 744 & 718  & 26  & 35 & 425&
375  &  7  &  8  & 78  & 76\\
  11   & 4  &  3 & 318 & 318  & 35  & 37 & 779 & 755  & 30  & 34 & 455&
409  &  5  &  6  & 83  & 82\\
  12   & 3  &  3 & 321 & 321  & 26  & 34 & 805 & 789  & 20  & 17 & 475&
426  &  4  &  4  & 87  & 86\\
  13   & 3  &  2 & 324 & 323  & 13  & 23 & 818 & 812  & 12  & 21 & 487&
447  &  5  &  4  & 92  & 90\\
  14   & 2  &  1 & 326 & 324  & 15  & 15 & 833 & 827  & 10  & 20 & 497&
467  &  7  &  2  & 99  & 92\\
  15   & 2  &  4 & 328 & 328  & 11  & 14 & 844 & 841  & 12  & 16 & 509&
483  &  5  &  4 & 104  & 96\\
\hline
\end{tabular}
\end{table*}

We feel the outcome of this test against the recent XMM1/ChaMP results
validates our techniques of likelihood calculation and {\it ROSAT} field
shifting.  Accordingly we present this whole-sky-based optical analysis
against the {\it ROSAT} catalogues as a best-effort bulk astrometric solution of
the {\it ROSAT} field positions.  Such an optimized statistical approach will
always contain individual errors of course, but we trust that our
generally correct results will aid future research which will over time
improve our knowledge of the details.

\subsection{The Radio Sources}

Unlike the X-ray catalogues, the radio catalogues (NVSS, FIRST and SUMSS)
do not take the approach that each detected object is a discrete source,
as extended emission and lobes are found as commonly as detections of
point-like objects.  Accordingly the only warning flag accompanying the
data is that of possible false detection, for example for such
observational artefacts as sidelobes of bright sources.  FIRST and SUMSS
provide such flags, and we do not use data bearing those flags.  NVSS is
already clean.

These radio surveys are astrometrically well-grounded and do not require
field shifting as did the X-ray surveys.  Early pre-publication versions
of this catalogue did detect and utilize some field shifting of the radio
catalogues, but further examination showed these shifts to be spurious and
based on coincidence.  In the end the only discrepant astrometry arises
from the APM raw astrometric offsets from POSS-I and UKST plates which are
up to 2 arcsec in RA and DEC, see MWHB for a full discussion.  Thus it is
our optical catalogue which diverges from the true, not the radio
catalogues.  But we had already taken the decision to use our optical
astrometry as master, so we needed to align the radio astrometry to the
APM astrometry, i.e. to shift the radio fields up to 2 arcsec in RA and
DEC where required, using the same likelihood algorithm as was used for
the {\it ROSAT} fields.  We have performed this adjustment on a field by field
basis which works well for the small FIRST fields but is less effective
for the large NVSS and SUMSS fields, for which the astrometric uncertainty
for detections is typically 2 arcsec anyway.  In practice a very few
fields shift as far as 3 arcsec in RA or DEC which we take as an
accumulation of astrometric and positional errors and rounding, corrected
by the shift.  With the astrometry aligned, we applied our likelihood
algorithm to detect core radio-optical associations.  Totals for the three
input catalogues are listed in Table 10, and field-by-field summaries can
be found at http://quasars.org/docs/radio-fields.zip .  Note the better
weight-per-association ratio for the FIRST detections compared to NVSS and
SUMSS which results entirely from the better astrometric fit to our
optical background.

\begin{table}
\caption{Numbers of radio detections and $>70$\% confidence radio-
optical source associations for each radio source catalogue.}

\begin{tabular}{lrrrr}
\hline
Source   & No. of   &   No. of radio& No. of $>$70\% core  &   Total\\
catalogue& fields  &  detections &  associations &   weight\\
\hline
FIRST   &   29148   &    781667  &      134444  &   121523.4\\
NVSS    &    2326    &  1810664 &       142268  &   106358.9\\
SUMSS   &     428  &     165531   &      27126  &    19442.5\\
\hline
\end{tabular}
\end{table}

Of course much of the significance and interest inherent in radio
detections is in identifying extended radio double lobes and associating
them with source optical objects.  But our all-sky-based likelihood method
is effective only for core detections.  The lobe detections as recorded in
the radio catalogues are typically offset too far away from the optical
source for our likelihood algorithm to confer more than a token
probability of association.  We found we needed to devise a heuristic
pattern analysis algorithm to identify lobe candidates, using in turn the
attributes of each of the three input radio catalogues.  Such pattern
analysis cannot be done from first principles.  In the radio catalogues
the detection entries have been reduced from raw data and formatted into
flux ellipses of specified axes and orientation angles.  Large lobes are
often represented as many ellipses, especially in the FIRST catalogue.
Our task is to find the rules which work best to identify these ellipses
as lobes where they are in fact lobes; identifications can ultimately be
confirmed on a case-by-case basis by comparison with images from the
surveys' respective image servers.  Of course, many images look
inconclusive.  If we find the rules which will reliably accord with the
conclusive cutout images, then we will be content with the algorithm's
judgement for the inconclusive images.  These heuristics should apply to
orthogonally significant aspects of the input catalogue data, whose
contributions to our overall confidence in each double-lobe identification
can be quantitatively assessed.

As ours is an optically-based catalogue we concerned ourselves only with
those radio-emitting objects which are detected on our optical catalogue.
Many bright lobes originate in objects too optically faint to appear here.
In such cases we are in danger of falsely attributing the lobes to a
nearby optical object.  The single clearest indicator of such a false
declaration is for the optical object to be offset from the natural
midline of the two lobes.  This can be described in terms of the angle
subtended by the two radio signatures about the optical object.  `Perfect'
lobes make a $180\degr$ angle with the optical centroid; angles less than this
are not uncommon as lobes bend in the IGM, so a lower angle can be valid,
but as the angle grows smaller it becomes more likely that we are simply
using the wrong optical centroid.  This angle of the lobes about the
presumed optical centroid is our first criterion for assessing candidate
lobes, and, as will be seen, it is also the determinant by which we
discern population excesses over the background which yields a total count
of double lobes for us to locate.  We have chosen to permit double lobe
configurations with a source bending angle (lobe-identification-lobe) of
$\ge 90^\circ$ only, which we expect will have little impact on
completeness.

We first collect the set of all double lobe candidates together with
candidate optical sources.  We treat as a lobe candidate any radio
detection which does not lie within 2 arcsec of an optical object.  Thus
if a true lobe happens to be at the same position as an unrelated optical
object we will both declare a false core association and exclude that lobe
from our search for double lobes!  Such errors are unavoidable, but such
precise chance alignment must be rare, and the fact that we have
encountered only a single instance of it in testing against known double
lobes persuades us that the problem is small.  We search the sky for
optical objects within 90 arcsec of every radio detection not already
associated with an optical source.  Every optical object thus picked up by
two separate radio detections becomes a candidate optical centroid
provided the angle subtended by the two radio detections about it is
$\ge 90^\circ$.  Of course in a field with many radio detections and optical
objects this can produce a great many permutations with many candidatures
for each object.  We need to find the best unique lobe candidates for an
optical centroid, and a best unique optical centroid for each lobe pair;
ideally, this should correctly correspond to the real lobes and their true
optical sources on the sky.

To achieve this we identified, as an initial step, distinct criteria which
test the joint hypothesis that two radio detections are in fact a lobe
pair and that a certain optical object is their true centroid.  To know
the number of true lobes in the part of the sky under consideration would
be a great help as we could then compare our resultant lobe count to the
known total to see how well we are doing; in practice, we obviously do not
know the number of true lobes.  But as a substitute we are able to
identify excess non-random configurations of sources overlaying the random
background which constitute a potential separate radio population, i.e.,
the double lobes.  The identification of this excess population and the
application of our selected criteria proceeded together in an iterative
process applied to our data pool of candidate lobes and optical centroids,
as described below.

Our seven primary criteria to identify radio lobe pairs and their optical
centroids are:

\begin{enumerate}
\item Angle ($\theta$): angle subtended by the two radio sources about the optical object.
\item Distpct ($\delta$): Comparative offsets of the two radio sources from the optical
  object; the smaller offset is expressed as a percentage of the larger
  offset.
\item SNRpct ($R$): Comparative flux strength of the two radio sources, expressed as
  signal-to-noise ratio; the smaller SNR is expressed as a percentage of
  the larger SNR.
\item SDratio ($S$): SNR-to-offset comparison, designed to exclude weak radio sources
  at large radio-optical offsets as we model that large lobes should be
  brighter than the small lobes visible at the faint limit of these
  surveys; proportional to the minimum SNR/offset$^2$.
\item CLA ($\psi$): Comparative lobe angle of the two radio ellipses, expressed in
  degrees.  This compares the respective offsets of the ellipse major-
  axis orientation to optical-to-radio direction for each of the two
  radio sources, so CLA=0 shows a perfect match of the two lobe ellipses,
  as when there is e.g., a $20\degr$ clockwise tilt of each ellipse axis
  compared with its direction to the centroid.  This gauges the
  morphological similarity where the two radio sources are distant from
  the optical centroid, typically when the only parts of the lobe visible
  are the surrounds of the bright `hotspots' at the end of the jets.
  It is intended to penalize random isolated detections which are 
unrelated to the candidate centroid and point in unrelated directions.
\item EA ($E$): Eccentricity alignment of the radio detections.  This is for when the
  radio ellipses represent well-defined lobes extending away from the
  centroid, and combines the eccentricity $e$ of the radio ellipse with its
  angle of alignment ($\phi_A$) to the optical centroid; $\phi_A=0$ means the
  ellipse's major axis points back to the optical centroid.  For small
  double-lobe angles (see (i)) the lobes must be significantly elongated
  towards the candidate optical identification to make the optical object
  a strong candidate.  EA is expressed as $e\times (4-\phi_A/10)^2$ for
  $\phi_A<40\degr$, using the lesser score of the two lobes.  If EA is high then
  CLA will be high too, but this is desirable, since we consider a high
  EA a strong lobe signature.
\item Offset ($\Delta$):  Large optical-radio offsets are more likely to be a consequence
  of random alignment, so we need to assign an offset-based penalty to
  keep these out, expressed as the mean of the two lobe optical-radio
  offsets, in arcsec.
\end{enumerate}

We next needed to assess the relative weight to be given to each of
these criteria; e.g., how much better is an angle of $180\degr$ than
140\degr, or, if all else is the same, how much better is it if the
radio-optical offset is 10 arcsec instead of 80 arcsec? This involved
an iterative analysis of the radio-optical data where in turn the
impact of varying weights for one criterion is measured while the
other six criteria are held fixed. We found that four iterations of
this process yielded adequate stability for all the criteria, as well
as a viable figure for the excess, i.e. the expected number of double
lobes to be found. Once in possession of this robust excess, we
re-initialized the iterative analysis holding the excess as a
constant, and so refined the weightings. As enumeration of the excess,
i.e. the lobe population, is such a useful process, we next describe
how this was carried out.

The population of excess, double-lobe candidate sources is estimated
separately for each input radio catalogue by finding the excess of
large- angle configurations of two radio sources about each optical
centroid. We derive these excesses by analysing all double lobe
candidate configurations within a 90 arcsec radius of optical objects,
summarized in Table 11 for all such candidate double lobes of angle
$>115\degr$, in $5\degr$ bins centred about the listed values, except
for the $180\degr$ bin which is half- width. The total count of double
lobe candidates (non-unique in that individual sources are re-used
across multiple configurations) is displayed, as well as two kinds of
backgrounds to be deducted, `static' and `geometric'. There is also a
column of `best unique' candidate lobes which are selected by the six
other quantified criteria (SNRpct, CLA, etc); these candidate lobes
are matched to a single optical candidate, without duplication.

\begin{figure*}
\begin{center}
\epsfxsize 14cm
\epsfbox{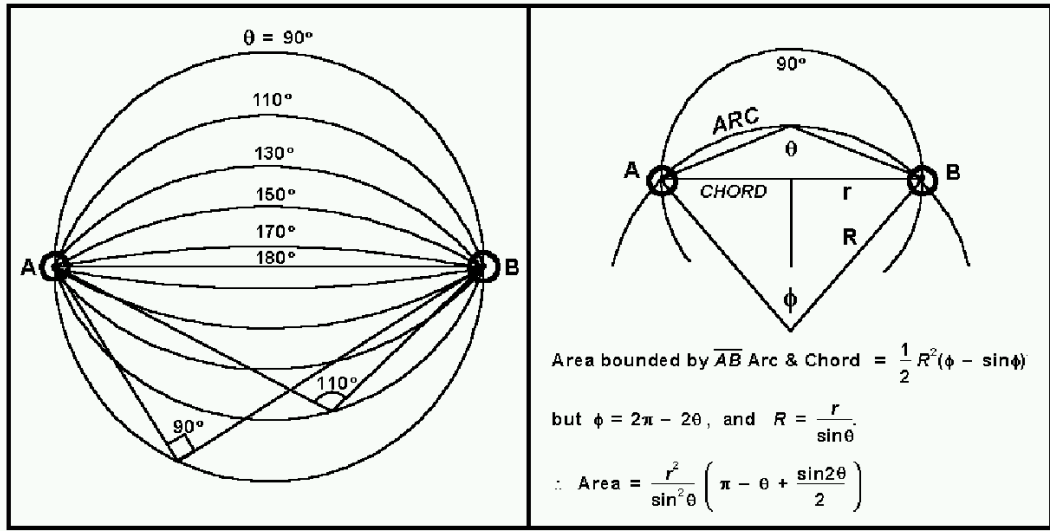}
\caption{The geometry of the double-lobe background calculation}
\end{center}
\label{geom}
\end{figure*}

The `static' background comes from considering random pairs of radio
detections around an optical object. We should of course expect to see
equal numbers of sources for all radio-optical-radio angles. This
background of random configurations dominates our set of candidate
lobes. The `geometric' background consists of false optical matches to
true double lobes, and is modelled by considering random optical
objects within a disk bounded by two lobes at opposite ends, which
contains the entire space for all angles $\ge 90^\circ$; Fig.\ \ref{geom}a
shows that each angle ($\theta$) space follows an arc which passes
through both lobes A and B, and Fig.\ \ref{geom}b shows that the arc is of a
notional circle of radius $R = r/sin\theta$ where r is half the
distance between the two lobes. An individual angle space (e.g.
$\theta=163\degr$) can be quantified by using the area of the segment
of disk R bounded by the arc and chord connecting the two lobes, as
shown in Fig. \ref{geom}b (where $\theta$ is expressed in radians); the angle
space area of e.g. $163\degr$ is the difference in areas of the disk
segments defined by $\theta=162.5\degr$ and $\theta=163.5\degr$. In
this way it can be shown that the normalized expectation of finding a
false lobe configuration per individual angle (e.g. 163\degr) ranges
from 200\% of mean at $90\degr$ to 66.67\% of mean at
$180\degr$, and has the form

\begin{eqnarray}
 e(\theta) = 180&\times&\left[{{1-(\theta-.5)/180+\sin(2(\theta-.5))/2}\over{\sin^2(\theta-.5)}}\right.\cr
&&            -\left.{{1-(\theta+.5)/180+\sin(2(\theta+.5))/2}\over{
     \sin^2(\theta+.5)}}\right]\cr
&&       (90\degr \le \theta  <
179.5\degr) 
\end{eqnarray}

The level of the geometric background cannot be estimated in isolation
as it depends on the presence of all true lobes, including those for
which the true optical identification is too faint to be found in our
optical catalogue. We choose to combine estimates of the static and
geometric background in such a way as to yield an angle-based excess
corresponding to our expectation that there will be few lobes with
angles of $<140^\circ$, with lobes increasing as we approach 180\degr.
The decisive constraint is that the excess lobe population should be
small and flat between $110\degr$ and $140\degr$, so we find static and
geometric populations which will match that expectation. In practice
this constraint imposes a delicate balance between the two
populations, and we consequently find that per single- angle bin
(e.g., 167\degr) of the FIRST lobe candidates the static background
value is about 6475 objects and the geometric background has a
coefficient multiplier of about 500 for equation (4). Similarly, 17000
and 2700 are the static and geometric per-angle values found which
suit NVSS, and 1300 and 215 are found for SUMSS. These
values, accumulated into $5\degr$ angle bins, yield the excess above
background shown in Table 11.

\begin{table*}
\caption{Double-lobe excesses derived for the three radio catalogues,
binned by $5\degr$ angles (180\degr\ is half-width). Columns are the
`double lobe candidates' based on all the permutations of double-radio
and single- optical configurations within disks of sky of 90 arcsec
radius, the static random background and geometric background derived
from our fit coefficients, and the residual excess after background
subtraction. `Best unique' lobe numbers are reductions by the
quantified criteria (CLA etc), yielding unique candidate counts for
each $5\degr$ bin. The values of `Ratio used' are generated by the angle
formulae from Table 12.}
\begin{center}
\begin{tabular}{lrrrrrrrr}
\hline
         &  &    Double lobe & Static & Geometric &            &Best &  Excess/ & Ratio\\
Survey & Angle (\degr)&  candidates  & background& background&   Excess &  unique& unique &    used\\
\hline
FIRST  &  115  &  35323  &   32375  &   2868   &   80  &  4985  & 0.016&
0.008\\
FIRST  &  120  &  35081  &   32375  &   2638   &   68  &  5210  & 0.013&
0.010\\
FIRST  &  125  &  34892  &   32375  &   2445   &   72  &  5482  & 0.013&
0.014\\
FIRST  &  130  &  34581  &   32375  &   2282  &   -76  &  5710  & $-$0.013&
0.019\\
FIRST  &  135  &  34784  &   32375  &   2146  &   263  &  6029  & 0.044&
0.027\\
FIRST  &  140  &  34646  &   32375  &   2035  &   236  &  6333  & 0.037&
0.036\\
FIRST  &  145  &  34659  &   32375  &   1940  &   344  &  6467  & 0.053&
0.050\\
FIRST  &  150  &  34447  &   32375  &   1862  &   210  &  6870  & 0.031&
0.069\\
FIRST  &  155  &  34984  &   32375  &   1801  &   808  &  7234  & 0.112&
0.094\\
FIRST  &  160  &  35211  &   32375  &   1751  &  1085  &  7607  & 0.143&
0.129\\
FIRST  &  165  &  35588  &   32375  &   1713  &  1500  &  8134  & 0.184&
0.176\\
FIRST  &  170  &  36185  &   32375  &   1687  &  2123  &  8456  & 0.251&
0.242\\
FIRST  &  175  &  37842  &   32375  &   1672  &  3795  &  9299  & 0.408&
0.331\\
FIRST  &  180(hw)& 19051  &   16188   &   750  &  2113  &  4773  & 0.443&
0.453\\
NVSS  &   115  &  99657  &   85000  &  15492  &  -835  & 10066  & $-$0.083&
0.003\\
NVSS  &   120  &  98872  &   85000  &  14242  &  -370  & 10791  & $-0.034$&
0.004\\
NVSS  &   125  &  98290  &   85000  &  13198   &   92  & 11232  & 0.008&
0.005\\
NVSS  &   130  &  97377  &   85000  &  12325   &   52  & 11914  & 0.004&
0.007\\
NVSS  &   135  &  96507  &   85000  &  11594  &   -87  & 12542  & $-0.007$&
0.010\\
NVSS  &   140  &  96060  &   85000  &  10983   &   77  & 13268  & 0.006&
0.014\\
NVSS  &   145  &  95824  &   85000  &  10475  &   349  & 13764  & 0.025&
0.020\\
NVSS  &   150  &  95999  &   85000  &  10057  &   942  & 14509  & 0.065&
0.029\\
NVSS  &   155  &  95607  &   85000  &   9722  &   885  & 15277  & 0.058&
0.041\\
NVSS  &   160  &  95804  &   85000  &   9454  &  1350  & 15923  & 0.085&
0.057\\
NVSS  &   165  &  95820  &   85000  &   9254  &  1566  & 16640  & 0.094&
0.081\\
NVSS  &   170  &  96127  &   85000  &   9113  &  2014  & 17486  & 0.115&
0.115\\
NVSS  &   175  &  96830  &   85000  &   9031  &  2799  & 18011  & 0.155&
0.162\\
NVSS  &   180(hw)& 48587  &   42500  &   4050  &  2037  &  9382  & 0.217&
0.230\\
SUMSS  &  115  &   7756   &   6500  &   1234   &   22  &   579  & 0.038&
0.003\\
SUMSS  &  120  &   7550   &   6500  &   1134  &   -84  &   655  & $-0.128$&
0.005\\
SUMSS  &  125  &   7566   &   6500  &   1050   &   16  &   737  & 0.022&
0.007\\
SUMSS  &  130  &   7569   &   6500   &   981   &   88  &   811  & 0.109&
0.010\\
SUMSS  &  135  &   7428   &   6500   &   923   &    5  &   850  & 0.006&
0.014\\
SUMSS  &  140  &   7412   &   6500   &   875   &   37  &   887  & 0.042&
0.020\\
SUMSS  &  145  &   7488   &   6500   &   834  &   154  &  1053  & 0.146&
0.028\\
SUMSS  &  150  &   7339   &   6500   &   801   &   38  &  1073  & 0.035&
0.039\\
SUMSS  &  155  &   7305   &   6500   &   775   &   30  &  1117  & 0.027&
0.056\\
SUMSS  &  160  &   7430   &   6500   &   753  &   177  &  1186  & 0.149&
0.079\\
SUMSS  &  165  &   7535   &   6500   &   736  &   299  &  1322  & 0.226&
0.111\\
SUMSS  &  170  &   7375   &   6500   &   725  &   150  &  1395  & 0.108&
0.157\\
SUMSS  &  175  &   7549   &   6500   &   719  &   330  &  1466  & 0.225&
0.222\\
SUMSS  &  180(hw)&  3754   &   3250   &   322  &   182  &   763  & 0.239&
0.314\\
\hline
\end{tabular}
\end{center}
\end{table*}

Table 11 shows the static and geometric background figures that we
subtract from the total number of possible candidate lobes to yield
the excess of lobe candidates above the background expectation. The
excess totals to about 12000 for FIRST, 12000 for NVSS and 1500 for
SUMSS, so these are approximately the numbers we'll be trying to
locate. The next column, `best unique', gives the number of best
unique lobes identified by the lobe selection criteria. Our task is to
find which of these best unique candidates are the genuine lobes
enumerated by the excess. The next column `excess/unique' shows the
fraction of the best unique candidates that we expect will be genuine
lobes. The last column is the per-angle value of the formula (shown in
Table 12) that we design to simulate the ratio. This formula is
applied to the angle of each candidate double lobe as an absolute
starting point; thus, for example, a FIRST candidate with angle of
152\degr\ is assigned the angle-based expectation of 0.0777, since we
nominally expect 7.77\% of these objects to be true lobes. Our final
score for each double-lobe candidate will be on an open-ended scale
normalized to a score of 1 equating to a 50\% probability of being a
lobe, i.e. score = odds/(1-odds). So we first convert our angle-based
expectation into that scale, so 0.0777 becomes 0.0777/(1-0.0777) =
0.0843. For this candidate now to attain a final 50\% score it will
need to gain a total multiplier of 11.9 from the other six selection
criteria all of which have had curve-fitting formulae similarly
constructed as above, but are normalized to 1 equating to 50\%.
These formulae for the seven quantified criteria, displayed in Table
12, are only heuristic data-curve fitters and do not have any physical
meaning. The importance of fitting curves closely was brought home to
us when we initially derived curve-fits for only the FIRST catalogue
and applied them to the NVSS: results were sparse. So we have elected
to prefer exactitude over simplicity in designing these formulae, but
we emphasise that they are just heuristic estimators.

\begin{table*}
\caption{Formulae used to calculate quantitative criteria to evaluate
candidate double lobe configurations.}
\begin{center}
\begin{tabular}{lllll}
\hline
Criterion & FIRST formula    &     NVSS formula   &        SUMSS formula&
Notes\\
\hline
Angle ($\theta$)      &$2^{(\theta-156)/11}/10$& $2^{(\theta-158)/10}/20$ & $2^{(\theta-
155)/10)}/18$\\
Distpct ($\delta$)    &$1+(\delta-65)/90$   &  $1+(\delta-62)/62$    &  $1+(\delta-62)/62$\\
SNRpct    ($R$) &$2^{(R/45)-1}$  &     $2^{(R-62)/12}$  &   $2^{(R-75)/12}$&eq 0.5 if $S=100$ (sidelobe)\\
SDratio ($S$) &  $2^{\log_2(S)-6}$ &  $3^{\log_2(S)-5.2}$ & $4^{\log_2(S)-
4.2}$& max 35\\
CLA    ($\psi$) &   $5^{(1-\psi/35)}$     &     $4.2-\psi/9$      &        $3.5-\psi/15$&
min 0.15, max 3 for FIRST\\
EA   ($E$)  &    $1.75+E/2$        &     $1.33+E/1.33$      &     $0.5+E/2$&
0.33 if $E=0$, max 25\\
Offset ($\Delta$)  &  $2^{(31-\Delta)/6}$   &  $2^{(46-\Delta)/12}$  &   $2^{(54-\Delta)/6}$&min 0.1, max 1\\
\hline
\end{tabular}
\end{center}
\end{table*}

As an example of how these formulae were derived, the data which yielded
the CLA formulae are presented in Table 13.  `Best' candidates are
compared to all candidates.  A `best' candidate is one which scores above
the mean for each of all the other criteria, and so is double-lobe-like in
every way.  We conjectured that these well-behaved candidates were true
lobes and so used them as a control population (we confirmed that more
than 95\% of them were likely lobe detections by inspection of
images from the surveys).  We derived the ratio of these best candidates
to all the candidates for each CLA value binned by 5\degr, and normalized this
about the mean.  The last column for each survey is the formula-derived
score, using that survey's CLA formula from Table 12 which replicates the
normalised ratio.  Thus the cumulative effect of applying these scores to
the data is that the total score is approximately unchanged.

\begin{table*}
\caption{Comparative Lobe Angle dependent attributes of the input radio
catalogues.}
\begin{tabular}{lrrrrrrrrrrrr}
\hline
  &   FIRST  & FIRST &  normal  &  FIRST &  NVSS  &  NVSS  &normal  &  NVSS&
SUMSS &  SUMSS & normal &  SUMSS\\
CLA  &  best &  backgd  & ratio  &  calc &  best &  backgd  &ratio  &  calc&
best & backgd &  ratio &   calc\\
\hline
 0(hw) &  90  & 17403  & 3.57  &  3.00  &  45  & 49684 &  4.94  & 4.20
&5  & 2356  & 3.17  & 3.50\\
 5  &   153  & 35087  & 3.01  &  3.00  &  74  & 99889 &  4.04  & 3.64
&9  & 4665  & 2.88  & 3.17\\
10  &   137  & 34753  & 2.72  &  3.00  &  62  & 98794 &  3.42  & 3.09
&17  & 4754  & 5.34  & 2.83\\
15  &   128  & 34921  & 2.53  &  2.51  &  45  & 98635 &  2.49  & 2.53
&4  & 4711  & 1.27  & 2.50\\
20  &   104  & 34733  & 2.07  &  1.99  &  28  & 98228 &  1.56  & 1.98
&5  & 4700  & 1.59  & 2.17\\
25  &   85  & 34547  & 1.70  &  1.58  &  33  & 98125 &  1.84  & 1.42
&5  & 4683  & 1.59  & 1.83\\
30  &   54  & 34543  & 1.08  &  1.26  &  16  & 98174 &  0.89  & 0.87
&6  & 4714  & 1.90  & 1.50\\
35  &   48  & 34397  & 0.96  &  1.00  &   5  & 98194 &  0.28  & 0.31
&3  & 4778  & 0.94  & 1.17\\
40  &   26  & 34640  & 0.52  &  0.79  &   3  & 98105 &  0.17  & 0.15
&1  & 4707  & 0.32  & 0.83\\
45  &   31  & 34368  & 0.62  &  0.63  &   5  & 98206 &  0.28  & 0.15
&1  & 4722  & 0.32  & 0.50\\
50  &   17  & 34572  & 0.34  &  0.50  &   3  & 97806 &  0.17  & 0.15
&1  & 4539  & 0.32  & 0.17\\
55  &   12  & 34624  & 0.24  &  0.40  &   4  & 98020 &  0.22  & 0.15
&.  & 4871  &  .  &   0.15\\
60  &   14  & 34687  & 0.28  &  0.32  &   2  & 98461 &  0.11  & 0.15
&.  & 4750  &  .  &   0.15\\
65  &    5  & 34828  & 0.10  &  0.25  &   .  & 98511  & .  &   0.15
&.  & 4651  &  .  &   0.15\\
70    &  .  & 34867    &  .  &  0.20  &   .  & 98254  & .  &   0.15
&.  & 4729  &  .  &   0.15\\
75    &  .  & 34813    &  .  &  0.16  &   .  & 99331  & .  &   0.15
&.  & 4817  &  .  &   0.15\\
80    &  .  & 34685    &  .  &  0.15  &   .  & 98455  & .  &   0.15
&.  & 4717  &  .  &   0.15\\
85    &  .  & 34616    &  .  &  0.15  &   .  & 99380  & .  &   0.15
&.  & 4805  &  .  &   0.15\\
90(hw)  & .  & 17257    &  .  &  0.15  &   .  & 49372  & .  &   0.15
&.  & 2456  &  .  &   0.15\\
\hline
\end{tabular}
\end{table*}

Each double-lobe candidate is scored using the seven criteria, and the
individual scores are multiplied together to give the total normalized
score for the candidate, so a total score of 1 indicates that the
candidate is about 50\% likely to be a lobe.  Since the starting
score from the angle-based excess is of the order of 0.1, it is clear that
a lobe candidate will need to pick up good scores from a number of these
criteria to achieve a high score, signifying a true lobe.  We also use two
additional normalized criteria which aid in choosing an optical source for
a double lobe where there are multiple optical candidates: (1) Core radio
detection:  An optical candidate directly detected in radio is a very
strong source candidate for suitably configured lobes.  We have quantified
this as a 15x multiplier via analysis against `best' candidates similar to
that presented for CLA, above.  (2) Optical morphology and colour:  This
quantifies which types of optical objects are most likely to be associated
with radio lobes, gauged again by analysis of the data as with CLA.
Objects absent in one colour are only one-third as likely to be radio
emitters, and objects stellar in red are two-thirds the likelihood.
Objects that are non-stellar in red, i.e. galaxies, are 2.25 times as
likely to be the core object.  Blue colour morphology is weighted as for
the red colours but with half the significance.  Colours (B-R) impact the
final likelihood in a range from .33x to 3.5x; in the case of stellar
objects it is the blueish objects which are favourable and the reddish
unfavourable, whilst with galaxies it is the reverse.  These two centroid-
based multipliers are removed after de-duplication, so do not contribute
to the final score on which the lobe-ness of the candidate is assessed.
One artefact which caused us some trouble was that some sidelobes still
remain unflagged in the source catalogues; these appear as regularly-
spaced spikes ringing bright sources and so score quite well on some of
our tests, but their very regular nature allows us to trap and remove them
with some success, as with perfectly matched SNRpcts on Table 12 where we
assign a low score. We also removed pairs with very faint SNRs where the
ellipses were perfectly round -- this too denoted sidelobes.

When all two-radio-one-optical candidate scoring has been completed, all
candidates scoring less than 33\% are discarded and the rest are de-
duplicated by peeling off the top; that is, accepting the top-scoring
combinations and then removing any other candidates that were sharing
those radio or optical objects, and repeating to completion.  Thus we are
left with completely unique two-radio-one-optical candidates with final
probability scores.  To clarify the status of low-scoring candidates we
found it useful to apply our standard likelihood algorithm treating the
lobes as highly-offset core detections; the average density score at high
offsets is 1 (= background), but for some optical PSFs and colours it is
greater than 1 and for others less, so the likelihood algorithm confers an
additional judgement on whether that class of centroid typically shows
large-offset associations above the background, i.e., lobes.  Thus our
last step to these lobe probability calculations is to add the likelihood
density to the lobe probability score and treat this final score as a
likelihood density figure so that a density of 2 equates to a confidence
of 50\% using equation (2), etc.  We apply a cutoff at
confidence=40\% and the surviving double-lobe candidates are
accepted for inclusion into our catalogue.  Comparison of our results with
images from the radio surveys shows our results to be in good agreement
with the radio images, i.e. where our catalogue says we find lobes, they
generally do look like lobes.

To check our results more stringently, we compare them with a radio
survey with pre-existing optical identification, the online 3CRR
catalogue (Laing, Riley \& Longair 1983) at
http://www.3crr.dyndns.org/ . Because this is a low-frequency flux
density selected sample, it contains the brightest lobes; as these are
often large and nearby, we do not expect to detect all of the lobes
given our size limit of 90 arcsec. However, there are few radio source
surveys with a high enough identification fraction to meet our needs.
We considered using the $z\sim 1$ B2/6C `Distant DRAGNs' survey (Eales
et al. 1997), in which the lobe sizes are generally smaller, but of
their 27 IR-detected centroids only two are bright enough in V to
appear in our optical catalogue! As it happens, both centroids
(0901+35 and 1045+35A) have double lobes of LAS $<8$ arcsec so that
FIRST reports them as single detections only and so appear in our
catalogue as core-detected objects QORG J090432.3+353904 and QORG
J104830.4+353801. Of Eales' remaining double lobes, one pair (0905+39)
is declared in QORG associated at 85\% confidence to a nearby
(23 arcsec) false centroid QORG J090818.8+394319, and the remainder
are excluded due to the absence of any suitable optical objects. This
result gives a sense of the optical faintness of $z\sim 1$ galaxies
which are not optically selected, and is encouraging in the sense that
there was only one QORG assignation of double lobes to a false
centroid where 14 double lobes (and 12 core detections) are seen in
the FIRST data.

Returning to the 3CRR catalogue, it lists 173 optical identifications,
of which 13 are core detections only, for which inspection of
FIRST/NVSS reveals 80 core (offset $\le 3$ arcsec) radio detections
and $\sim 90$ possible lobe pairs within 90 arcsec of the centroids.
However only 132 of the 173 centroids appear in our optical data
copositioned within 8 arcsec of the listed 3CRR position, and of
these, 10 have no core radio detection nor more than one FIRST/NVSS
radio signature within 90 arcsec. Of the remaining 122 optical
centroids we find our catalogue has associated 62 to core radio
detections; 8 additional core detections were rejected by our
likelihood algorithm as they are astrometrically offset too far from
the optical centroids. Also for these 122 optical 3CRR centroids,
inspection of FIRST/NVSS reveals $\sim 70$ possible lobe pairs within
90 arcsec from which our QORG algorithm identified 58 lobe pairs and
associated 43 of these with optical objects that we here find are the
correct 3CRR centroids for a 74\% hit rate on double- lobe
declarations; about 6 more were associated with optical signatures
within 20 arcsec of the 3CRR centroids which look possibly related.
Given that FIRST tends to break these large, bright lobes down into
multiple ellipses, we feel that our algorithms have performed
reasonably here. In all, 102 out of the eligible 122 optical 3CRR
centroids are radio- associated in our catalogue. The complete list of
the 3CRR centroids and our results for them is viewable at
http://quasars.org/docs/3CRR-QORG.txt, which also displays confidence
percentages for those near-core radio- optical superpositions which
were rejected. In the QORG catalogue we have retained, for
consistency, those objects that this exercise has shown to be false
centroids for 3CRR double lobes, but we have annotated some of these
as `vicinity of' a 3CRR centroid to identify the lobes to the user.
This is a nod to our difficulty with lobes too large for our 90-arcsec
selection criterion; we found no lobe candidates at all for the 8 3CRR
galaxies listed with lobes of LAS$>1000$. However, such large sources
are likely to be comparatively rare. The large bulk of QORG
double-lobe declarations are for smaller lobes for which the centroid
identification is usually straightforward, except where the true
identification is too faint for our optical data, which is always a
hazard. Table 14 gives a summary of lobe counts binned by angular size
of the longer lobe.

\begin{table*}
\caption{Summary of double radio lobe numbers in the QORG
catalogue, by source catalogue and binned by angular size of the longer
lobe in arcsec.  Numbers of lobes, core detections, median lobe flux in
mJy and median confidence of the QORG lobe declarations rounded to 1 per
cent are displayed.}
\begin{tabular}{lrrrrrrrrrrrr}
\hline
Ang size  & \multicolumn{4}{c}{FIRST}& \multicolumn{4}{c}{NVSS} &\multicolumn{4}{c}{SUMSS}\\
longer & No.&  No. core&  median &  median  & No.  &No. core & median  & median& No. & No. core & median&   median\\
 lobe &   double &  det'ns&  flux&  conf &   double &  det'ns&  flux&  conf&   double &  det'ns&  flux&  conf\\
(arcsec)&lobes&&(mJy)&(per cent)&lobes&&(mJy)&(per cent)&lobes&&(mJy)&(per cent)\\
\hline
 2- 5  &   996  &     .  &   5  &   90  &      3  &    .  &   14  &   64&
1  &   .  &   12  &   63\\
 6-10  &  3142  &   196  &   5  &   96  &     19  &    1  &    6  &   63&
3  &   .  &  111  &   67\\
11-15  &  2278  &   546  &   5  &   93  &     30  &    1  &    5  &   84&
6  &   .  &  158  &   74\\
16-20  &  1531  &   385  &   5  &   93  &     37  &    .  &   10  &   77&
9  &   .  &  255  &   92\\
21-25  &  1082  &   255  &   6  &   90  &     88  &    4  &   20  &   89&
25  &   .  &   62  &   95\\
26-30  &   783  &   213  &   7  &   88  &    374  &   32  &   29  &   94&
117  &   .  &   35  &   94\\
31-35  &   562  &   110  &   7  &   84  &    756  &   55  &   31  &   97&
189  &   .  &   39  &   90\\
36-40  &   419  &    66  &   7  &   81  &    988  &   59  &   30  &   95&
244  &   .  &   46  &   91\\
41-45  &   268  &    53  &   7  &   77  &   1032  &   53  &   28  &   92&
237  &   .  &   40  &   90\\
46-50  &   160  &    30  &   8  &   75  &   1018  &   65  &   28  &   89&
196  &   1  &   44  &   89\\
51-55  &   117  &    10  &  10  &   68  &    915  &   46  &   29  &   87&
203  &   .  &   41  &   86\\
56-60  &    70  &     8  &   7  &   66  &    781  &   46  &   29  &   83&
137  &   1  &   40  &   83\\
61-65  &    42  &     8  &  11  &   66  &    628  &   36  &   29  &   85&
113  &   .  &   48  &   84\\
66-70  &    27  &     2  &  14  &   58  &    484  &   33  &   33  &   82&
80  &   .  &   55  &   79\\
71-75  &    11  &     1  &   8  &   58  &    383  &   20  &   34  &   80&
42  &   .  &   62  &   77\\
76-80  &    13  &     2  &  31  &   56  &    325  &   21  &   42  &   78&
29  &   .  &   77  &   76\\
81-85  &     4  &     1  &  21  &   58  &    252  &    9  &   45  &   77&
24  &   .  &   65  &   77\\
86-90  &     7  &     1  &  10  &   73  &    210  &    7  &   41  &   76&
8  &   .  &   63  &   70\\
\hline
TOTAL  &  11512  &  1887  &   6  &   91  &   8323  &  488  &   30  &   88&
1663  &   2  &   45  &   87\\
\hline
\end{tabular}
\end{table*}

Table 14 shows that the 5-arcsec-resolution FIRST detections yield increasing
lobe numbers at shorter angular size as expected from the increase in the
background population with greater distance.  NVSS and SUMSS have 45-arcsec
resolution so at smaller angular sizes there is an increasing tendency for
lobes to be merged into a single detection.  Our total double lobe counts
are seen to compare well with the calculated excesses from Table 11
(unsurprisingly, since those excesses provided our starting likelihood
probabilities).  For each individual double lobe candidate in the
catalogue we list the nominal confidence percentage that it is a true
double lobe with the stated optical centroid.  Of course sometimes we
select the wrong centroid, as seen in some 3CRR examples above, and some
of our double lobe declarations are in fact unassociated and not double.
There is necessarily some relation between our declared confidences and
actual performance in discerning true lobes but in the absence of a large
control sample we can only surmise that the relation is not too greatly
skewed.  Our total count of double lobe declarations is 21,498, of which
about half are rated with a confidence over 90\%, slightly over
half for FIRST and slightly under half for NVSS \& SUMSS.

In the end, the merit of our heuristic pattern detection algorithm for
double radio lobes about optical centroids is weighed by its performance
against the real sky.  For the difficult, large 3CRR lobes we have
achieved an accuracy of 74 cent per from a completeness of about 85 per
cent against the FIRST/NVSS data.  We expect better performance for the
larger population of smaller lobes.  We feel that our list of double lobe
candidates over the whole sky, while not constituting a fully identified
sample, gives the largest currently available sample of {\it probable} lobe
identifications, and as such will be a useful resource for future
research.

\subsection{Use of the Identification Catalogues}

The aim of our work has been to associate radio/X-ray detections with
optical objects. The identification of an optical object as a known
quasar, galaxy or star is important but secondary, in the sense that
we do not wish to have to assess the level of our confidence that we
have selected the correct optical signature. Accordingly we assign the
identification only where we are essentially certain of it, which
usually means astrometric alignment within 4 arcsec. The general
method is that for each input identification catalogue we analyse
offset annuli from the catalogued object positions to determine to
what radius an optical population is found which is over twice the
background; this is typically 4 arcsec for astrometrically accurate,
recent catalogues. In most cases only a single optical object is found
in our input catalogues within that radius, which we take as
unambiguous identification; if no object is found then that
identification is lost. Where there is more than one optical object
within the radius we are content to use them all from then on, in the
expectation that final selection of the correct object will come via
one of them being found to be associated with a radio/X-ray source;
where there is no radio/X-ray association the identification will not
be used in our catalogue anyway. We modify these criteria when suited
to a particular catalogue: for PGC galaxies we find that optical
identification out to a 30 arcsec offset from the catalogued position
is merited if the PSF is non-stellar, and for white dwarfs we find a
maximum 15 arcsec offset if the PSF is stellar, as some of its data
come from early surveys. Many identifications of course appear in many
different catalogues, often with different names, so we have elected
to use the earliest names available; thus we prefer galaxy names as
given in the PGC as these are historical in nature. Where the PGC does
not name a galaxy we use the earliest available name from another
catalogue, and galaxies present only as a nameless entry in the PGC we
write as e.g. PGC 12345, using the PGC number which is used by LEDA as
an unchanging identifier. In the case of redshifts we wish to use the
latest measured values as these are most likely to be accurate. Thus
an identification in our catalogue will often get its name from one
source and its redshift from another. Attributions for the names and
redshifts are displayed in our catalogue for each identification, with
the attribution references listed in the readme file; we give
references only to source catalogues, and those catalogues should be
consulted for information on the original identification. 

Where an optical object is claimed by both a galaxy catalogue and a star
catalogue we have set quality standards to decide between these, e.g. a
recent redshift confirms an object as non-stellar, or a stellar PSF
confirms a star identification over a galaxy identification without a
redshift, etc.  We have prepared a list of `interesting' dual star/non-
star identifications at http://quasars.org/docs/Star-NonStar-Duplicates.txt
  In this list we also flag when an object appears in our catalogue; it
is only for those objects that our choice of the correct identification is
important.  In use of the QORG catalogue it is important to bear
in mind that many such `stars' have been misclassified, especially those
from Tycho, HDx and GCVS which are not spectroscopically supported, and
bright Tycho stars may conceal the actual sources of radio/X-ray emission.
Thus any stellar identification, not already well-understood, that is
reported in our catalogue to be associated with a radio/X-ray detection,
should be considered suspect, especially for those few that are nominally
associated with double radio lobes. 

A key identification in this catalogue is that of catalogued QSOs and BL
Lacs; we identify these with optical objects even in the absence of a
radio/X-ray association.  We do this since QSOs are such significant
objects, and since they are all likely to be X-ray and (in the case of BL
Lacs at least) radio emitters whether we have detected them or not  We are
content to rely on the Veron catalogue as the arbitrator of QSO
identification, so we exclude objects identified elsewhere (e.g. CfA) as
QSOs which are absent from Veron, unless included by radio/X-ray
association.  Our faith in the judiciousness of the Veron catalogue is
partly prompted by that one of us (EF) has assisted in tidying up problem
areas in its recent releases, so we have some personal knowledge of its
strengths.  We endeavour to optically locate QSOs however practical.  Most
QSOs, especially the large number recently identified in SDSS and 2QZ
surveys, are unambiguously identified with isolated optical objects within
the usual 4-arcsec astrometric radius.  We have used these unambiguous QSO-
optical matches to construct a qualitative QSO optical profile which we
then apply to those cases where an identified QSO has multiple nearby
optical candidates; this plus magnitude comparison, plus a subset of
unambiguous {\it ROSAT}/radio detection locations, allows us to select a
superior optical candidate in nearly all cases.  As a tiebreaker between
two equally good candidates (which is rare) we simply select the nearer
one.  For a QSO listed with a magnitude near or below our plate depth, if
there is no faint optical candidate within 4 arcsec we discard that QSO as
being undetected.

A special case in QSO identification is that of the older QSO surveys
of the 1980s and 1990s as listed in the Veron catalogue. These are
often listed with significantly discrepant astrometry and photometry
making computerized identification problematic. We have found that the
discrepancies often appear to have a systematic component peculiar to
each original survey. Thus we find, for each original survey, the
astrometric and photometric offsets to our optical catalogue for those
(often few) unambiguous identifications, and then applying those
offsets to all that survey's QSOs and then re-matching to our optical
catalogue, repeating in an iterative process until stability is
reached. This can result in unambiguous recovery of many or most of
the optical objects matching those QSOs, and we also use the
above-mentioned qualitative profile to find most-likely optical
candidates offset up to 40 arcsec. In this way we have assisted our
recovery of about 200 QSO-optical identifications for the old surveys,
confirmed by comparison of a large subset to the original finding
charts, which we accordingly include in QORG with an astrometric and
photometric accuracy not found elsewhere. Surveys thus given
interesting shifts are displayed at
http://quasars.org/docs/Personal-Equation.txt, although we leave off
those QSOs that were subsequently re- surveyed, for which updated
information is available in the latest version of the Veron catalogue.

Table 15 summarizes the identification catalogues contributing to
QORG. The CfA Redshift catalogue is itself a compendium of many
catalogues and papers, and includes the main LBQS and LCRS data
leaving us to add in those residual star identifications separately.
The CfA, NED, White Dwarf, PGC and Veron catalogues are collections of
heterogeneous data which have been standardized somewhat by those
catalogues' authors whilst retaining historical names; the other
catalogues are more internally consistent and often derived from
single surveys. Use of name and positional information directly from
the large SDSS and 2QZ catalogues allows consistent presentation
across different object types, which we prefer over the short forms
used in the Veron catalogue. Where we use the name of an object we
also use its type (quasar, galaxy etc.) supplied by that catalogue
except that we use any Veron-supplied type. Many catalogues categorize
galaxies into subtypes like NELGs, but such distinctions are unclear
for many galaxies and heterogeneously applied across catalogues, so we
thought it cleaner to simply defer to the Veron categorization of some
galaxies as AGN and leave the rest annotated just as `galaxies'. Thus
we show just five identification types in QORG: 49743 galaxies, 48285
quasars, 14633 AGN, 6314 stars and 841 BL Lacs. There are also 91
objects listed as `U' for unidentified, where a redshift or other
information is displayed. We include into QORG all QSOs, AGN and BL
Lacs for which we find an optical object; the others require a
radio/X-ray association for inclusion. Note that the LBQS stars data
contain one BL Lac identified in the original paper which was
inadvertently left off their catalogue (P.\ Hewett, private
communication); it is included in our catalogue.

\begin{table}
\caption{Identification catalogues. See the `source catalogues'
section for full attributions. Types of objects identified in each
catalogue are listed in order of their numerical prevalence: Q=QSO,
A=AGN, B=BL Lac, G=galaxy, S=star, U=unknown. The total used for names
in QORG includes all identified objects plus 91 unknown objects
bearing names.}
\begin{center}
\begin{tabular}{lrrrr}
\hline
          &  Object&   Total no.& No. used& No. used\\
          &  types &   unique  & for name  & redshift\\
Catalogue  &  incl  &  objects  & in QORG   & in QORG\\
\hline
2dFGRS  &     GS  &   236078  &   3403  &   2916\\
2QZ  &      QSABGU  &  40439  &  22077  &  22019\\
3CRR  &      GQA  &      173  &     49  &     43\\
6dF     &     G  &     15035  &    663  &    853\\
6QZ  &      SQABG  &    1529  &    261  &    265\\
CfA  &      GQABS  &  234703  &   2564  &   6432\\
Common Names & S  &      1127  &    173  &      --\\
CV     &      S  &      1143  &    184  &      --\\
ENEAR  &      G  &      1174  &     12  &     25\\
GCVS     &    S  &     10553  &    146  &      --\\
HDx     &     S  &     88831  &    200  &      --\\
LBQS stars &   SB  &     1390  &      1  &      --\\
LCRS stars  &  S     &    886  &      2  &      --\\
NED  &      (all)  &  (lots)  &     52  &     52\\
NLTT     &    S  &     71663  &    235  &      --\\
PGC (LEDA) &   G  &   1088795  &  38611  &   1250\\
PSCz     &    GS  &    15423  &    301  &    811\\
SDSS  &      GSQU  &  181959  &  23282  &  20209\\
Tycho  &      S  &   2539737  &   4871  &      --\\
UGC     &     G  &     13390  &     37  &    267\\
Veron  &    QABSGU  &  64942  &  22404  &  27536\\
White Dwarfs&  S  &      2206  &     97  &      --\\
Yale     &    S  &      3131  &    204  &      --\\
Zwicky  &     G  &     19372  &     78  &   2958\\
\hline
TOTAL  &    SGQABU     &   &     119907  &  85636\\
\hline
\end{tabular}
\end{center}
\end{table}

In all, we have tried to include all computer-processable identifications
extant in the literature to provide a fully annotated picture of the known
radio/X-ray ({\it ROSAT}) sky.  We use these identifications to calculate odds
that unidentified radio/X-ray objects are in turn quasars, galaxies or
stars, as explained in the main section of this paper.

\subsection{Attributes of this Catalogue}
\label{attributes}

\subsubsection{De-duplication and identification}

After construction of the catalogue we found it necessary to perform some
de-duplication because of large bright objects such as plate-saturating
stars or large galaxies with multiple components which manifest as
multiple optical signatures; these attract associations from multiple
{\it ROSAT} fields where in fact both X-ray sources and optical signatures are
just duplicates.  To allow this situation to go uncorrected would diminish
the ease of use of the catalogue and possibly mislead the user.  About
1500 such duplicates across different {\it ROSAT} fields have been removed or
amalgamated via preferential retention of associations to the bright
central star or galaxy, while closely adjacent associations within the
same {\it ROSAT} field are preserved.  The radio surveys have a separate issue
that resolved FIRST double lobes are often presented by the NVSS as a
single central source, which would constitute a false core detection if
left unattended; we have removed the NVSS association in those cases,
which number about 750.  These de-duplications have clarified cases of
multiple associations across radio and X-ray catalogues, and condensed our
catalogue by about 0.5\%.

We have made an adjustment to the likelihood-of-association probabilities as one
of the last acts of writing this catalogue. Small numbers variations
in the density calculations have occasionally resulted in large
densities at up to 30 arcsec offset, and at large offsets it is also
common to encounter multiple optical candidates which would decrease
the odds of association for any one of them. We have attached an
additional likelihood penalty to far-offset associations to take
account of the increased presence of multiple candidates. We used a
simple rule of thumb for offsets greater than 6 arcsec, subtracting
1/6 density point for each additional arcsec offset, e.g. at 9 arcsec
2.3 becomes 1.8. This dampens high-offset densities to where 21 arcsec
is the furthest offset for any $>70$\% confidence association
presented, and 26 arcsec for any at all. To put this in perspective,
95\% of all our presented core associations are offset within 8
arcsec, and 75\% are within 4 arcsec. This may be a
conservative measure, but we feel it is more excusable to
under-represent true far-offset associations than it is to
over-represent false ones.

\subsubsection{Distribution on the sky}

\begin{figure*}
\epsfxsize 17.5cm
\epsfbox{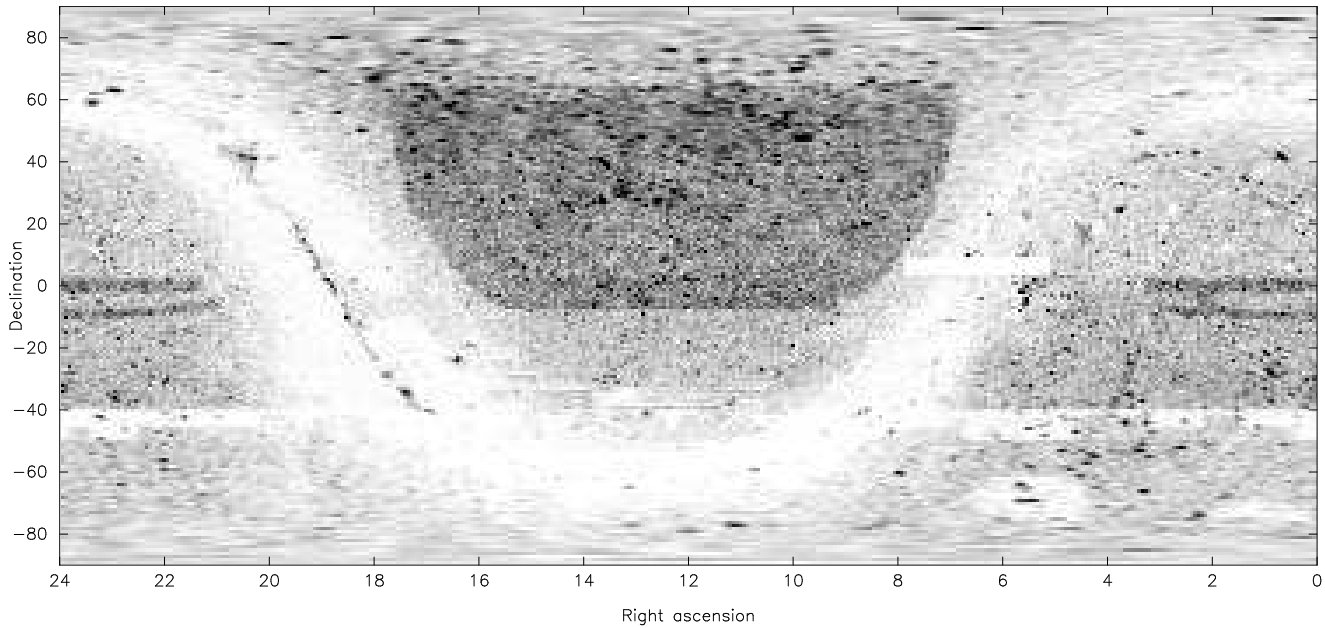}
\caption{A whole-sky optical density map (as Fig.\ \ref{qorg}) showing only
those 449,309 objects in our catalogue which are associated with
radio/X-ray detections.}
\label{rxray}
\end{figure*}

\begin{figure*}
\epsfxsize 17.5cm
\epsfbox{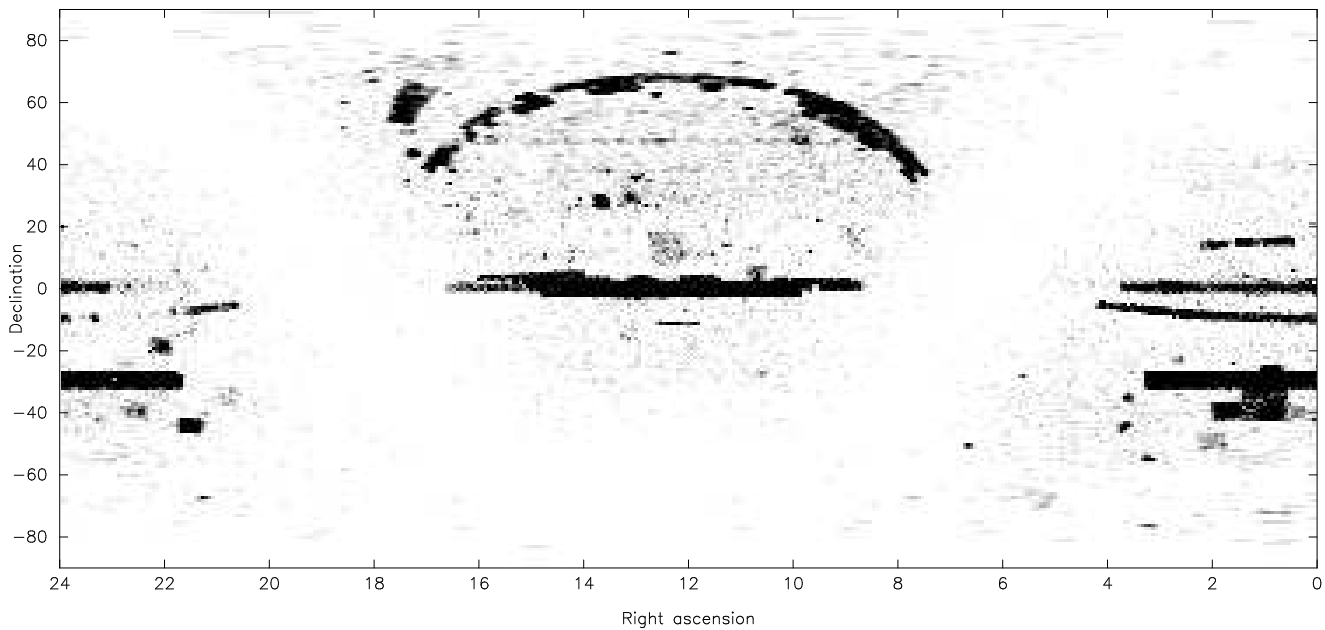}
\caption{A whole-sky optical density map (as Fig.\ \ref{qorg}) of the 53,930
catalogued QSOs (including BL Lacs and stellar-PSF AGN) found in our
catalogue.}
\label{qsos}
\end{figure*}

\begin{figure*}
\epsfxsize 17.5cm
\epsfbox{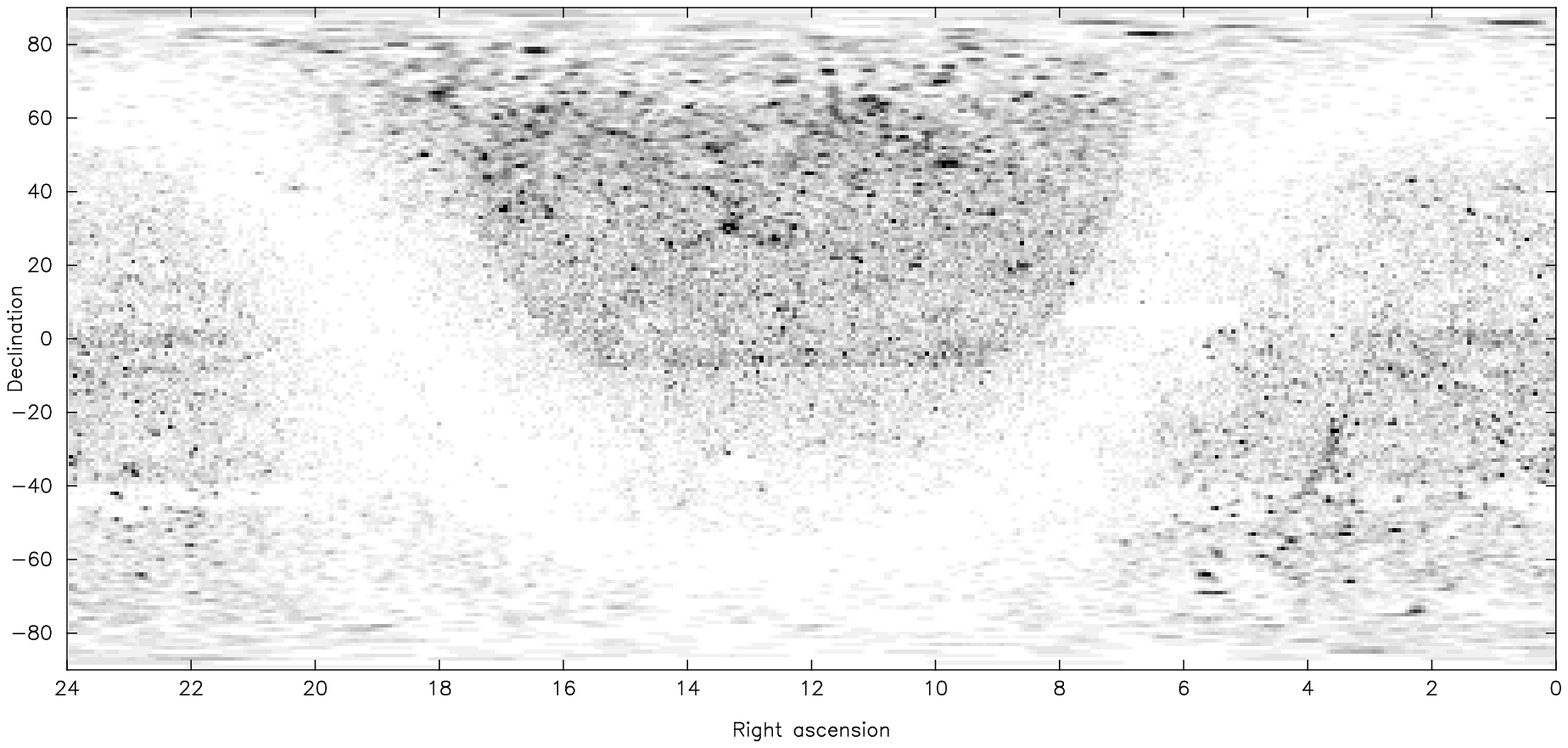}
\caption{A whole-sky optical density map (as Fig.\ \ref{qorg}) showing all 86,009 objects in our catalogue, not currently
identified, which we list as being 40\% to $>99$\% likely
to be a QSO.}
\label{qso-cand}
\end{figure*}

Because of the properties of the catalogues from which they originate,
the identified sources are not entirely uniformly distributed on the
sky. Fig.\ \ref{qorg} is a whole-sky optical density map of all
501,761 objects presented in the QORG catalogue. In the North
galactic cap (NGC) the large dense area in the centre is the FIRST
survey area, the dense equatorial strip is the part surveyed by both
SDSS and 2dF, and the crescent-shaped area to the North was surveyed
by the SDSS first release. In the SGC, the straight equatorial strip
and the curved strip below it are the FIRST south-sky survey area, the
dense straight area below that is 2dF-surveyed, and the white strip
below that is the part of the sky not surveyed in radio, with
NVSS-surveyed areas to the north and SUMSS to the south. The dense
strip in the Galactic dust lane to the East (left) shows an artefact
of our likelihood method where likely Galactic sources of radio/X-ray
emission are being presented as likely extragalactic; we have retained
these nominal associations in case some should prove useful, but users
are cautioned that probably most are spurious. In Fig.\ \ref{rxray} we
show a similar density map showing only those 449,309 objects in our
catalogue which are associated with radio/X-ray detections. The SDSS
and 2dF survey bands are missing from this map as they are
identification surveys, not radio/X-ray surveys. It can be seen that
the density of the radio/X-ray sources is quite uniform; the main
effect on the density, apart from the Galactic plane, is the area
covered by FIRST.

Turning to QSOs, Fig.\ \ref{qsos} is an all-sky density map of the 53,930
catalogued QSOs (including BL Lacs and stellar-PSF AGN) found in our
catalogue. It can be seen that 3/4 of them are concentrated into the recent
SDSS and 2QZ survey areas, and the remainder are inhomogeneously
distributed, showing how incomplete the overall QSO enumeration has
been to date. By contrast, Fig.\ \ref{qso-cand} is a similar density
map showing all 86,009 objects in our catalogue, not currently
identified, which we list as being 40\% to $>99$\% likely
to be a QSO. It can be seen that these are arrayed fairly uniformly on
the sky, barring the Galactic plane and the zone of declination $-45\degr$
which is as yet unsurveyed in radio.

\subsubsection{AGN properties}

\begin{figure*}
\begin{center}
\epsfxsize 12cm
\epsfbox{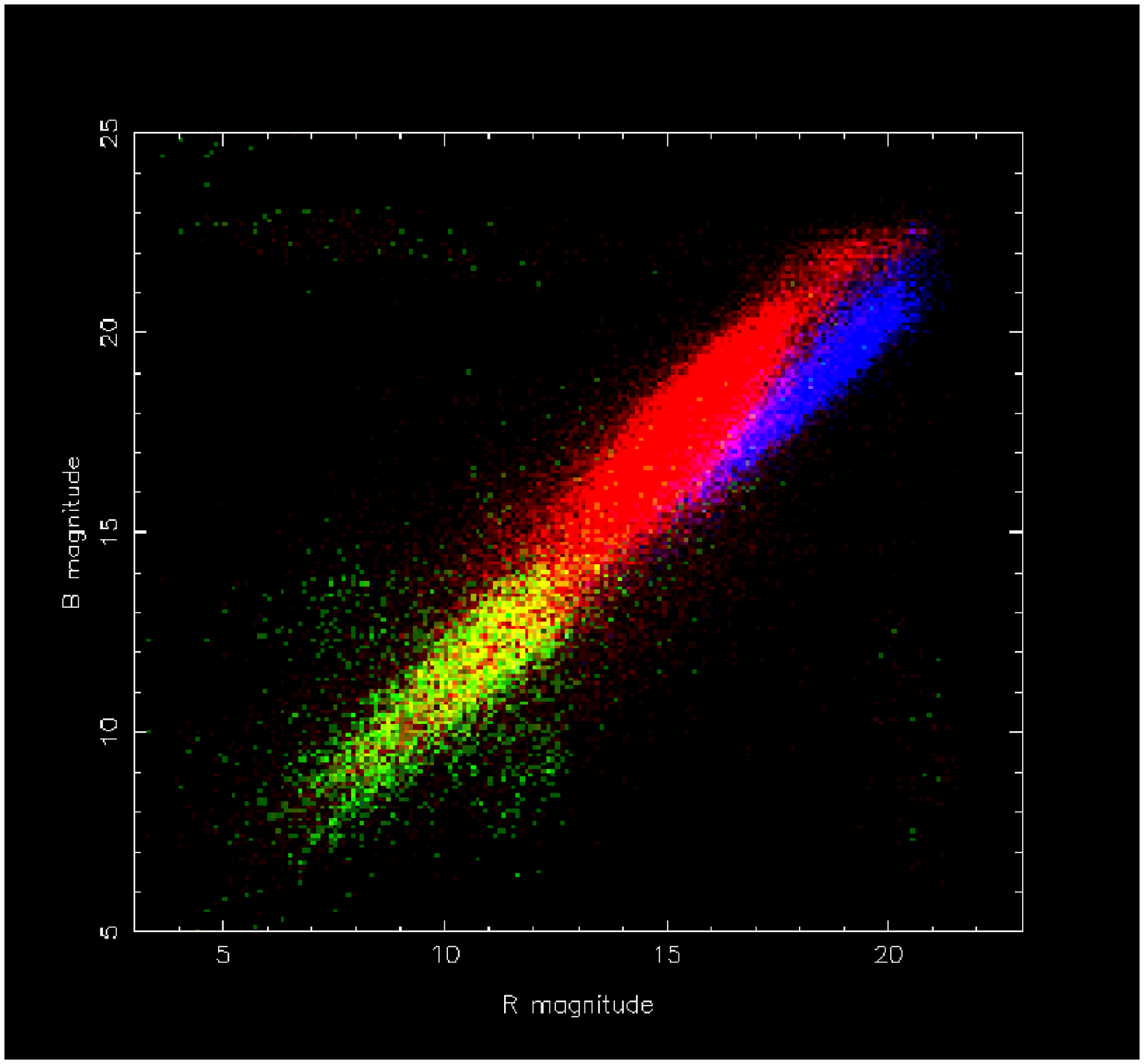}
\vskip 10pt
\epsfxsize 12cm
\epsfbox{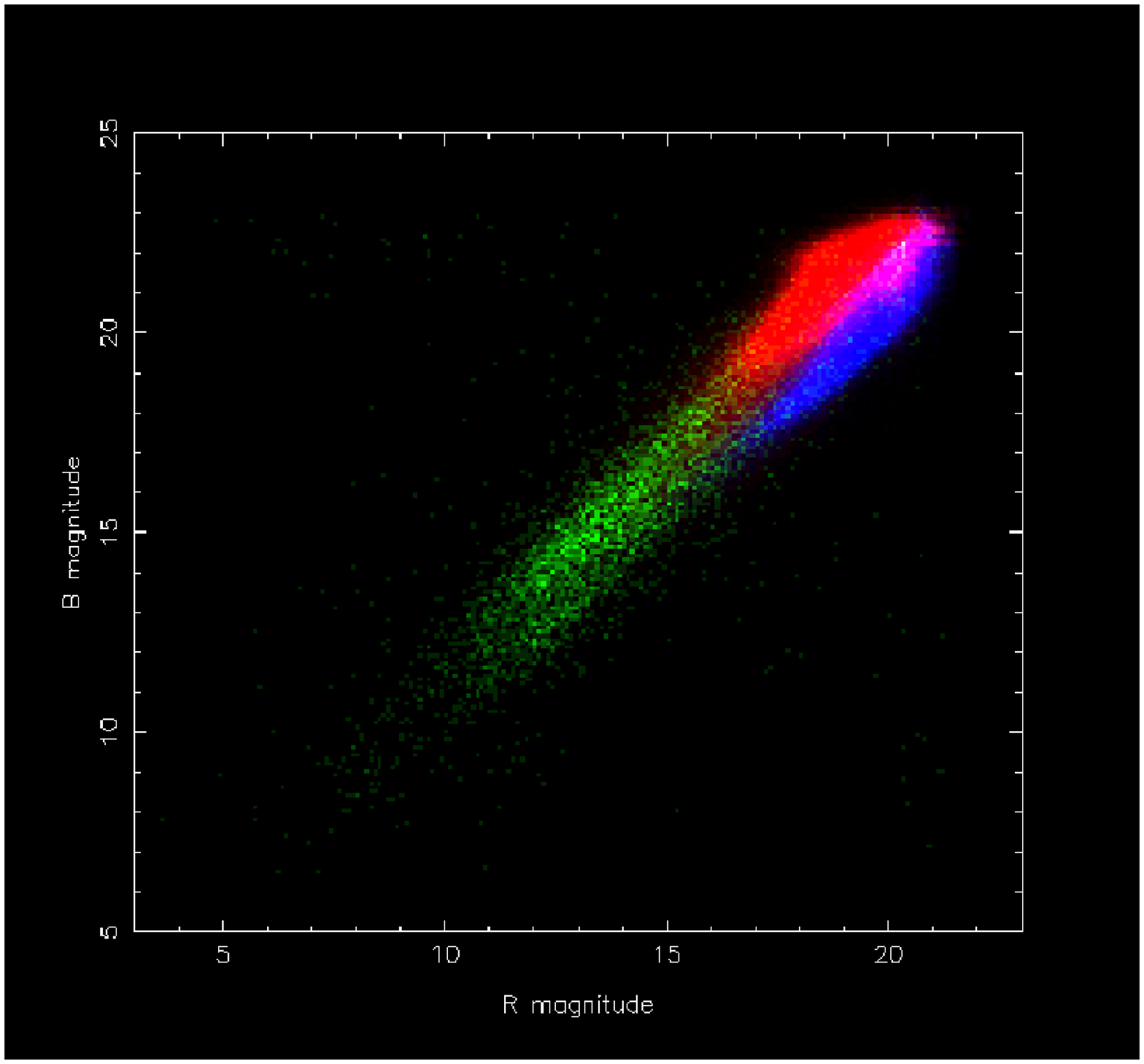}
\end{center}
\caption{$B$ against $R$ magnitude for (top) the previously identified
  sources and (bottom) the newly identified sources in the catalogue.
  Density of red, green and blue points represent density of sources
  identified in the catalogue as galaxies,
  stars and quasars respectively (only $>40$\% confidence
  identifications are used). Colours are additive in RGB colour space,
  so, for example, a magenta region on the plot represents a high
  density of both quasars and galaxies. Note that stars are
  over-represented for visibility.}
\label{colour}
\end{figure*}

We must necessarily begin this section of the discussion with a
caveat. Because the properties of the newly identified objects in our
catalogue are determined in a probabilistic way from the properties of
existing objects, it is dangerous to consider the statistics of the
newly identified objects and try to derive from them new results about
the population of X-ray and radio-identified optical objects as a
whole. For example, Fig.\ \ref{colour} shows plots of $R$ against $B$
magnitude for the previously identified and previously unidentified
sources in the sample, divided by object identification class. It will
be seen that the new sources lie in somewhat different areas of
parameter space (so, for example, there are few newly identified
galaxies with $R<15$, simply because the vast majority of resolved
objects with these magnitudes are already in catalogues). But the
important point is that the identification algorithm in general
populates a subset of the areas delineated by the existing data. It
cannot, by its nature, tell us more about the distribution of sources
with particular identifications in parameter space than the original
identification catalogue on which it was based.

\begin{figure*}
\begin{center}
\epsfxsize 12cm
\epsfbox{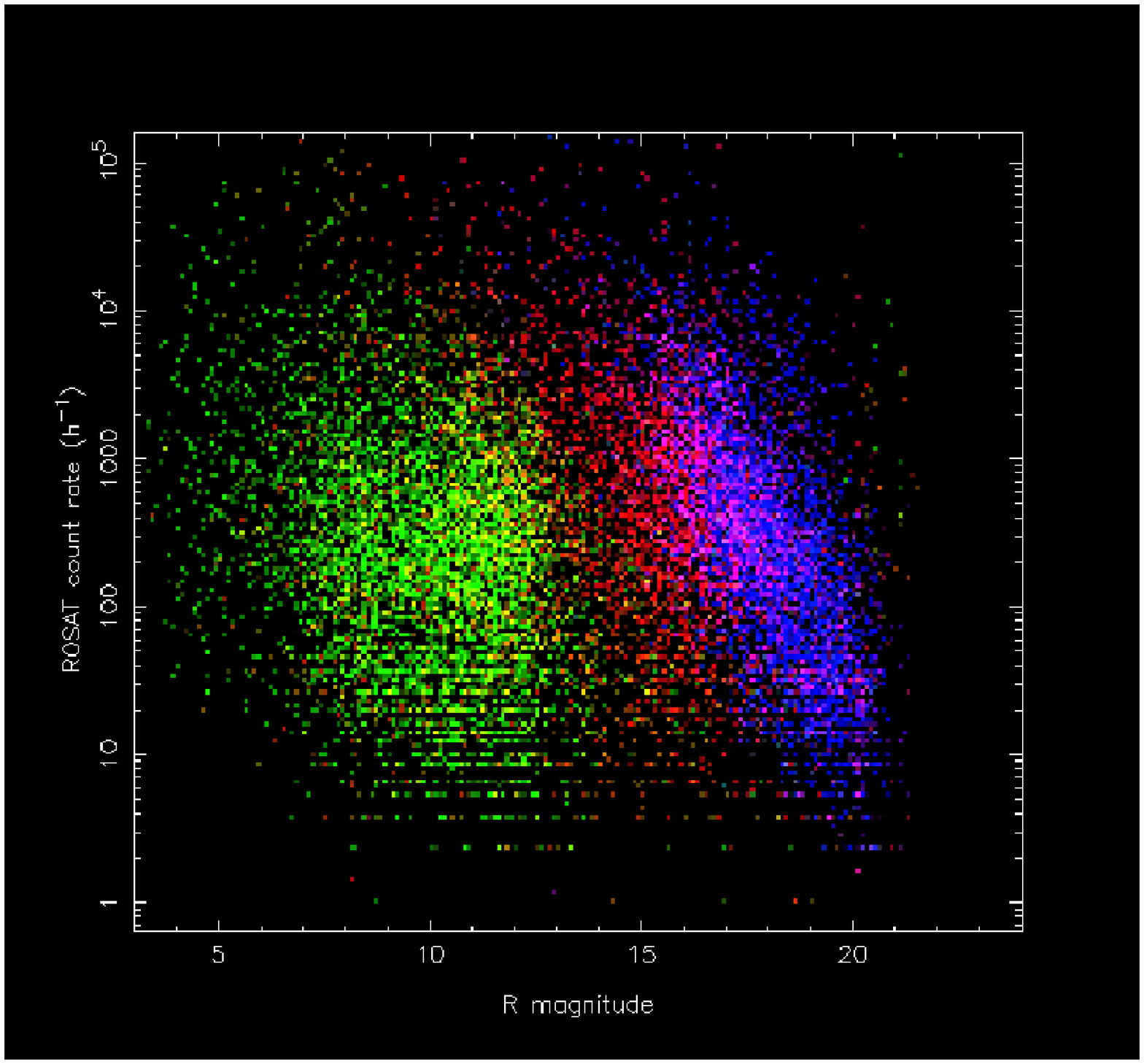}
\vskip 10pt
\epsfxsize 12cm
\epsfbox{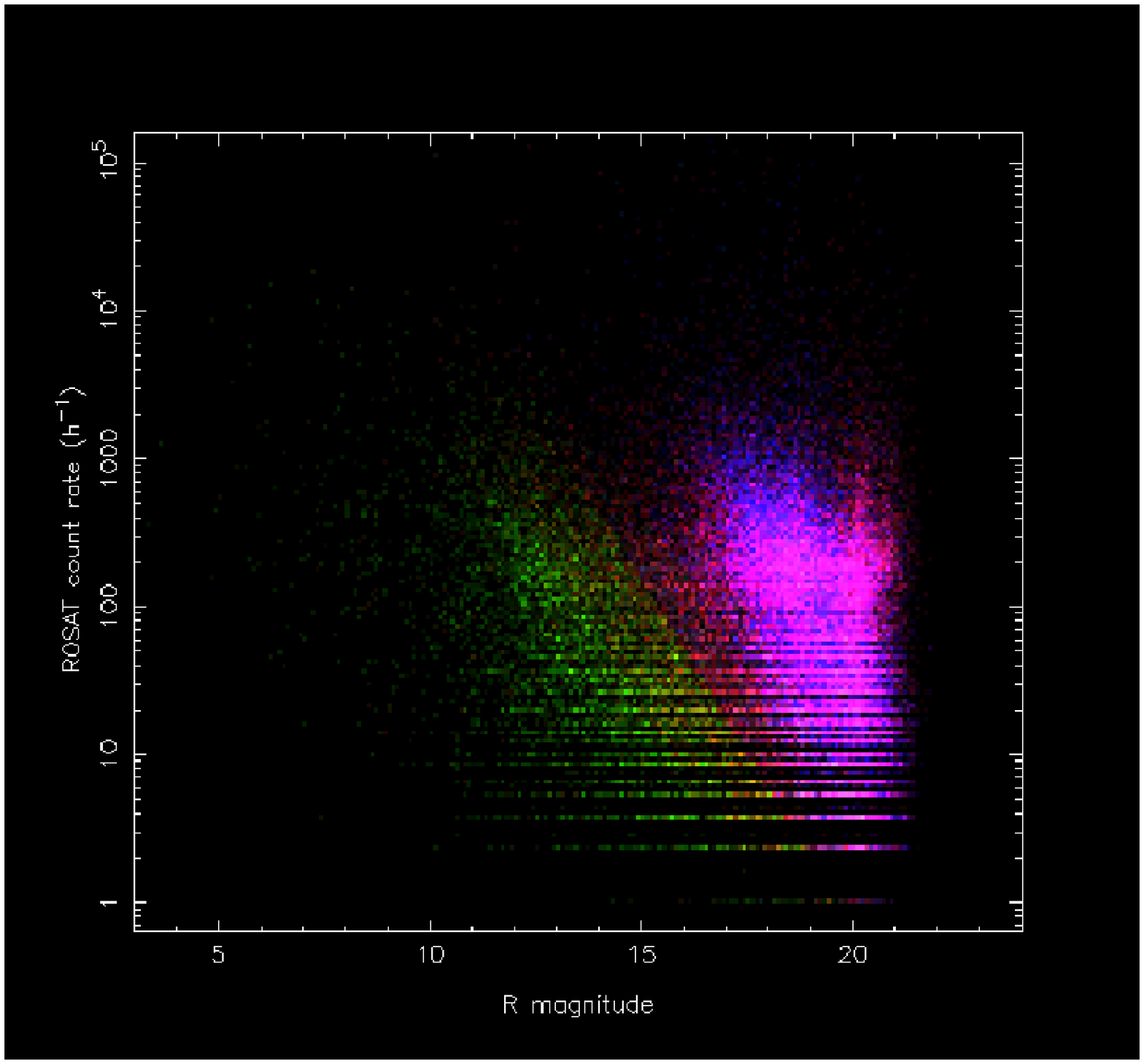}
\end{center}
\caption{{\it ROSAT} count rate (the mean of all available count
  rates, with the HRI value scaled up by a factor 3 to bring it in
  line with the PSPC values) against $R$ magnitude for (top) the
  previously identified sources and (bottom) the newly identified
  sources in the catalogue. Colours as for Fig.\ \ref{colour}. The top
  figure contains 13,733 data points, the bottom one 60,661.}
\label{xray}
\end{figure*}

With this in mind, it is worth carrying out a few simple analyses of
the characteristics of the objects in the catalogue. We begin by
examining the relationship between X-ray flux and optical magnitude
(Fig.\ \ref{xray}). The previously identified sources fall into clear
regions of parameter space; it is of course no surprise that for a
given X-ray count rate stars are generally optically brighter than galaxies and
galaxies brighter than quasars. The newly identified objects adopt
similar regions of parameter space, as expected, although there is a
relatively greater number of optically and X-ray faint objects. The
sharp line between galaxies and stars seen in the newly identified
objects is likely to be in part an artefact of the way that
radio/X-ray ratio is taken into account by the classification
algorithm (Section \ref{attributes}). At the optically faint end, the
probabilities that a given object is a quasar or a galaxy are similar
-- this reflects the difficulty in making a clear distinction between
the two types of object at faint magnitudes. We note that a {\it
  ROSAT} PSPC count rate of $\sim 10$ h$^{-1}$ corresponds to around
40 {\it Chandra} counts in 5 ks, and thus the types of sources being
identified here should be routinely found in {\it Chandra} and {\it XMM}
observations as (soft) serendipitous sources.

\begin{figure*}
\begin{center}
\epsfxsize 12cm
\epsfbox{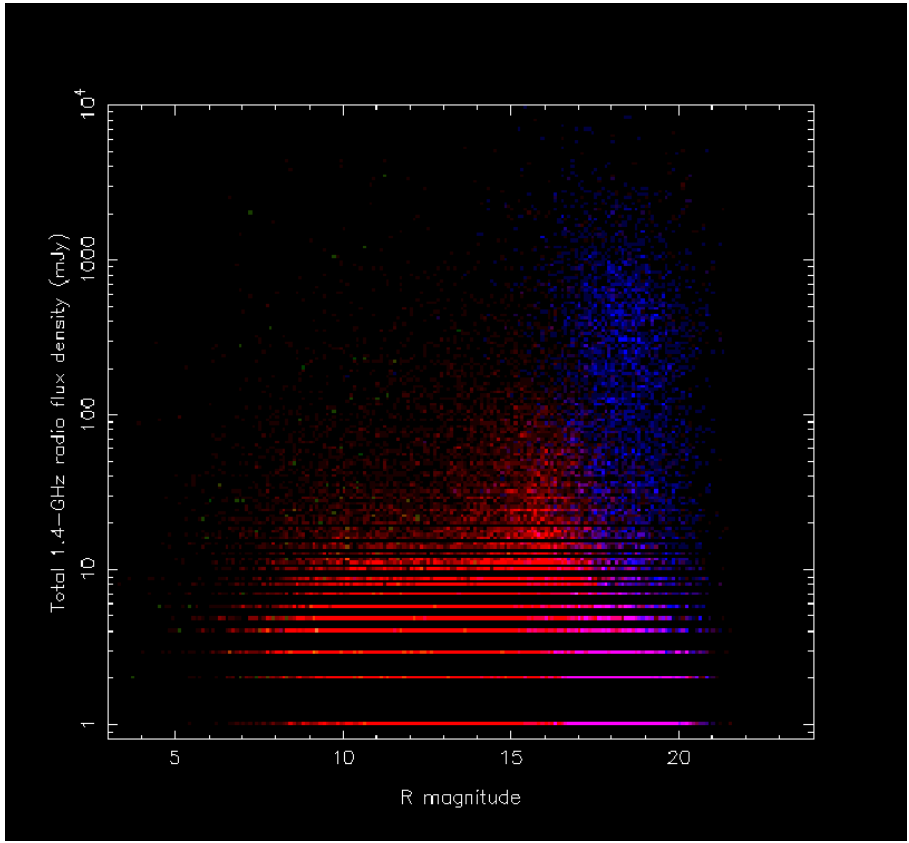}
\vskip 10pt
\epsfxsize 12cm
\epsfbox{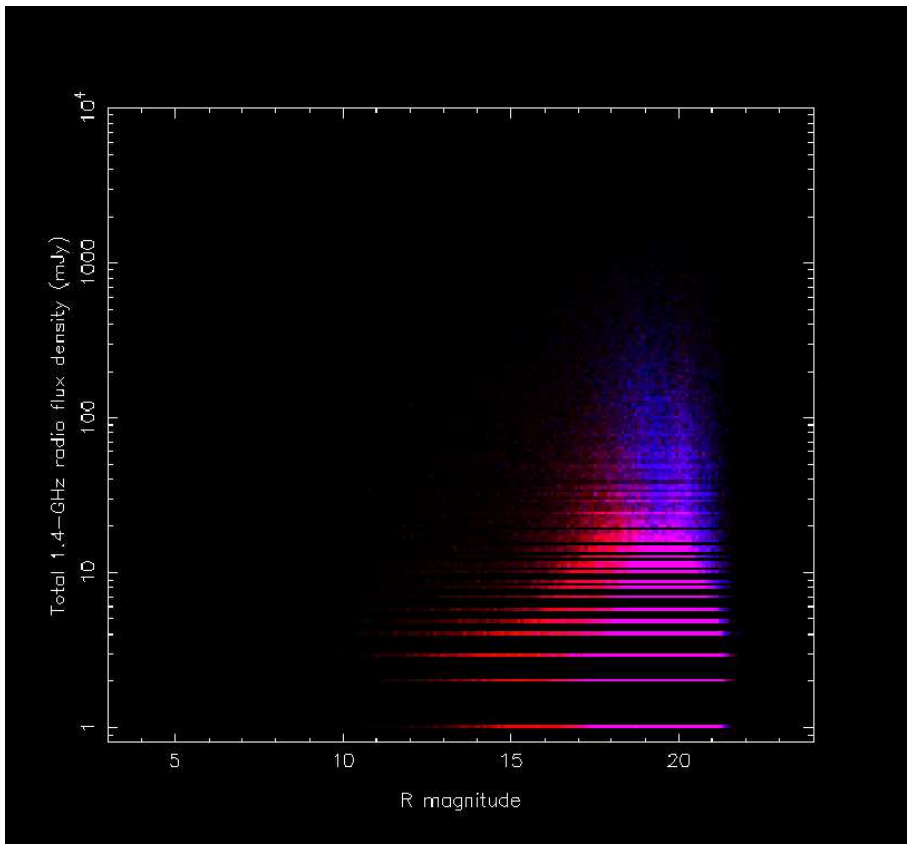}
\end{center}
\caption{1.4-GHz total flux density, including lobes where detected,
  from FIRST and NVSS, against $R$ magnitude for (top) the
  previously identified sources and (bottom) the newly identified
  sources in the catalogue. Colours as for Fig.\ \ref{colour}. The top
  figure contains 52,995 data points, the bottom one 274,505. Quasars
  are over-represented for visibility.}
\label{radio}
\end{figure*}

The corresponding radio plots (Fig.\ \ref{radio}) are also consistent
with expectation. Identifications with galaxies are most probable at
small magnitudes; quasars appear in large numbers above $R \sim 16$,
as seen in other catalogues, and above this magnitude the numbers of
galaxy and quasar identifications are similar, as expected from
unified models. As with the X-ray sources, the plot of newly
identified radio sources shows a higher density of galaxies at $R \sim
20$ than is seen in the identified sample. 

A small number of objects (5,325 galaxies and QSOs and a handful of
objects classified as stars, all at the $>40$\% level) are
identified as both radio and X-ray sources. While the QSOs show a
clear trend in the sense that X-ray count rate and radio flux density
are correlated, there is little evidence for a correlation between
these quantities for sources identified as probable galaxies. These
are likely to be more heterogeneous sources, including starbursts as
well as radio galaxies in a variety of environments. A trend in the
same sense is also present for sources with detected radio lobes and
X-ray counterparts, the vast majority of which are identified with
quasars.


\end{document}